\documentclass{jfm}

\usepackage{graphicx}
\usepackage{natbib}
\usepackage{upmath}
\usepackage{subfig}
\usepackage{booktabs}
\usepackage{afterpage}
\usepackage{amssymb}%
\usepackage{amsmath}
\usepackage{pgfplots}
\usepackage{amsbsy}
\usepackage{lineno}
\usepackage{etoolbox}

\allowdisplaybreaks

\newcommand*\linenomathpatch[1]{%
	\cspreto{#1}{\linenomath}%
	\cspreto{#1*}{\linenomath}%
	\cspreto{end#1}{\endlinenomath}%
	\cspreto{end#1*}{\endlinenomath}%
}

\linenomathpatch{equation}
\linenomathpatch{align}

\begin{document}
	\shorttitle{Kelvin wake pattern at small Froude numbers} 
	\shortauthor{R. Pethiyagoda, T. J. Moroney, C. J. Lustri and S. W. McCue} 
	
	\title{Kelvin wake pattern at small Froude numbers}
	
	\author
	{
		Ravindra Pethiyagoda\aff{1},
		Timothy J. Moroney\aff{1},
		Christopher J. Lustri\aff{2}
		\and
		Scott W. McCue\aff{1}\corresp{\email{scott.mccue@qut.edu.au}}\break
	}
	
	\affiliation
	{
		\aff{1}
		School of Mathematical Sciences, Queensland University of Technology, Brisbane QLD 4001, Australia
		\aff{2}
		Department of Mathematics, Macquarie University, Sydney NSW 2109, Australia
	}

	\maketitle
	
	\begin{abstract}
{ The surface gravity wave pattern that forms behind a steadily moving disturbance is well known to comprise divergent waves and transverse waves, contained within a distinctive V-shaped wake.  In this paper, we are concerned with a theoretical study of the limit of a slow-moving disturbance (small Froude numbers), for which the wake is dominated by transverse waves.  We consider three configurations: flow past a submerged source singularity, a submerged doublet, and a pressure distribution applied to the surface.  We treat the linearised version of these problems and use the method of stationary phase and exponential asymptotics to demonstrate that the apparent wake angle is less than the classical Kelvin angle and to quantify the decrease in apparent wake angle as the Froude number decreases.  These results complement a number of recent studies for sufficiently fast-moving disturbances (large Froude numbers) where the apparent wake angle has been also less than the classical Kelvin angle.  As well as shedding light on the wake angle, we also study the fully nonlinear problems for our three configurations under various limits to demonstrate the unique and interesting features of Kelvin wake patterns at small Froude numbers.}

		
	\end{abstract}
	
	\begin{keywords}
		surface gravity waves, wakes, waves/free-surface flows
	\end{keywords}
	
	\section{Introduction}
	\label{sec:intro}
	
	We are interested in studying three-dimensional steady free-surface flows that are caused by a disturbance moving at a constant speed.  This disturbance could be a pressure applied to the fluid surface or due to a submerged body.  In each case, the wave pattern that forms behind the disturbance is made up of divergent and transverse waves in the same way as a ship wake is.  We focus here on the low Froude number regime which arises when the disturbance is travelling sufficiently slowly.
	
	In part we are motivated by renewed recent interest in Kelvin ship wakes, initiated by \citet{rabaud13}, who noted that for sufficiently fast-moving ships (large enough Froude numbers), contrary to commonly held views, the wake angle that is observed behind a steadily moving ship (which we call the apparent wake angle, $\theta_\mathrm{app}$) is less than the well-known Kelvin angle $\theta_\mathrm{wedge}=\arcsin(1/3)$ (obtained from the linear dispersion relation).  \citet{rabaud13} provided a controversial explanation for this phenomenon by relying on an assumption that a ship hull does not create waves that are longer than its length.  This study was followed up by \citet{darmon14} who provided a rather different explanation, namely that the apparent wake angle is naturally provided by lining up the {points of maximum wave amplitude along} the divergent waves.  A theoretical analysis of the far-field wake behind a steadily moving pressure disturbance shows that this interpretation of apparent wake angle is consistent with the data of \citet{rabaud13} for large Froude numbers (see also \citet{verberck13,dias14} for a summary of these first two papers).  On the other hand, for real ships, there is an argument that the driving physics behind the reduced apparent wake angle is wave interference between divergent waves generated at the bow and stern of the vessel \citep{he14,noblesse14}.
	
	Since then there have been several subsequent papers which focus on non-axisymmetric simplified ship models and interference effects \citep{benzaquen14,ma16,moisy14b,miao15,noblesse16,zhang15a,zhu17,zhu18,Wu2019}, effects of shear and finite depth \citep{ellingsen14,pethiyagoda15,smeltzer17,zhu15,li16,li18} as well as the effects of viscosity \citep{Liang2019}.  While the details of these linear studies differ, a summary is that for moderate Froude numbers the apparent wake angle is roughly the same as the Kelvin angle, while for large Froude numbers the apparent wake angle decreases like the inverse Froude number { or inverse Froude number squared depending on the shape of the ship \citep{moisy14b}}.
	
	To complement the above research, we wish to revisit some of these ideas, but instead concentrate on the Kelvin wake pattern for small Froude numbers.  This regime is interesting because it {turns out} that, for sufficiently slow-moving ships, the apparent wake angle decreases as the Froude number decreases.  Thus, we have another regime in which the apparent wake angle is less than the Kelvin angle.  For sufficiently low Froude numbers, however, we are no longer free to use the {points of maximum wave amplitude} to define the apparent wake angle, since the {points of maximum wave amplitude} all lie on the centreline.  The reason for this different wave structure is that low-Froude number flows are dominated by transverse waves, not the divergent waves, and transverse waves decay away from the centreline.  As such, we borrow an idea from \citet{darmon14} and consider an arbitrary percentage of maximum height and associate the apparent wake angle with the wedge that aligns with this percentage.  { Note that analyses of far-field ship waves in the low-Froude-number regime are rare, an exception being the technical reports by \citet{noblesse86c,noblesse86a,noblesse86b} who used the low-Froude-number-limit to simplify an integral over the hull and perform a stationary phase approximation.}
		
Furthermore, { as well as considering the apparent wake angle},
we are also motivated to study the effects of nonlinearity \citep{soomere07}. Nonlinear free-surface flows at small Froude numbers are difficult to study in three dimensions since the wavelength decreases roughly like the square of the Froude number and the amplitude decreases exponentially in the small Froude number limit \citep{keller79}.  Thus there are computational issues that are in addition to the usual challenges of computing fully nonlinear solutions to this difficult three dimensional problem.  From the perspective of formal asymptotics, the exponentially small amplitudes mean that the waves appear beyond all orders of a traditional perturbation expansion in powers of Froude number squared \citep{lustri13}.  For this reason, there has been little analytical progress for the nonlinear version of this problem.  { One study by \citet{hermans89} used ray theory to evaluate wave profiles and resistance for weakly nonlinear ship waves, accounting for the near field water displacement by a hull.}
	
	In this paper, we consider three flow configurations: flow past an applied pressure distribution, which is often used as a simple model for flow due to a ship \citep[see][for example]{darmon14,ellingsen14,moisy14a,moisy14b,pethiyagoda15}; flow past a submerged source singularity \citep{forbes89,lustri13,pethiyagoda14a}, which can be thought of as a building block for three dimensional flows \citep{noblesse78,noblesse81}; and flow past a submerged doublet \citep{havelock32a}, which acts as a toy model for flow due to an underwater vessel \citep{arzhannikov16,kim69,pethiyagoda14b,scullen98}.  In \S\ref{sec:flowconfig} we lay out the governing equations for these flow configurations and provide a summary of the physical interpretation in each case.  We also explain the definitions of apparent wake angle that are used in this study.  Section~\ref{sec:linear} is devoted to the linear versions of our flow configurations which have the advantage of giving rise to known exact solutions.  We apply the method of stationary phase to derive explicit formulae for the apparent wake angle and use exponential asymptotics as a tool to propose an envelope function within which the apparent wake exists.  Our results show precisely how the wake narrows as the Froude number decreases.  In \S\ref{sec:slowlymovingobject} and \ref{sec:slowlymovingsourcedoublet} we treat two different small Froude number limits of the fully nonlinear problems of flow past a submerged source and doublet, while in \S\ref{sec:nonlinearflows} we undertake a numerical study in which we compute highly nonlinear solutions for small Froude numbers.  The wave patterns we observe in this regime appear to have interesting features that have not been recorded previously.  Finally, in \S\ref{sec:discussion} we summarise our results and discuss their significance.

	\section{Flow configurations and apparent wake angle}\label{sec:flowconfig}
	
	\subsection{Shared governing equations}
	
	For each of the three flow configurations considered in this paper, we assume the fluid is incompressible and inviscid and that the flow is irrotational.  We denote the location of the unknown free surface by $z=\zeta(x,y)$.  The velocity potential $\phi(x,y,z)$ therefore satisfies Laplace's equation throughout the flow field:
	\begin{equation}
	\nabla^2\phi=0,\qquad\text{for }z<\zeta(x,y).\label{eq:laplace}
	\end{equation}
	In dimensional units we suppose the disturbance (submerged source, submerged doublet or pressure) is moving with speed $U$ and is associated with a representative length scale $L$.  We set $g$ to be acceleration due to gravity. { We choose to ignore the effects of surface tension to specifically observe the effects of a low Froude number on gravity waves (as opposed to gravity-capillary waves). In a real world context our observations would apply to large cargo ships, near-surface submarines manoeuvring slowly, or slow moving currents past floating platforms, for example}.  Thus, by fixing our frame of reference to move with the disturbance, the dimensionless kinematic and dynamic conditions on the free surface are
	\begin{align}
	\phi_x\zeta_x+\phi_y\zeta_y&=\phi_z,\qquad\text{on }z=\zeta(x,y),\label{eq:kincond}\\
	(\phi_x^2+\phi_y^2+\phi_z^2)+\frac{2\zeta}{F^2}+\delta p(x,y)&=1,\qquad\text{on }z=\zeta(x,y),\label{eq:dyncond}
	\end{align}
	where the Froude number is defined by
	\begin{equation}
	F=\frac{U}{\sqrt{gL}}.
	\label{eq:Froude}
	\end{equation}
	The term $\delta p(x,y)$ is an applied pressure on the surface (in addition to atmospheric pressure), as described below.  The appropriate far-field conditions for our dimensionless problems are
	\begin{align}
	\phi&\sim x,\quad\zeta\rightarrow 0&\text{as }x\rightarrow-\infty,\label{eq:radiation}\\
	\phi&\sim x,\qquad&\text{as }z\rightarrow-\infty.\label{eq:farfield}
	\end{align}
	Equation (\ref{eq:radiation}) enforces the radiation condition that surface gravity waves do not propagate ahead of the disturbance, while (\ref{eq:farfield}) simply ensures that the flow approaches a uniform stream in the infinitely deep limit.

	\subsection{Flow past a submerged point source}
	\label{sec:flowsource}
	
	The first flow configuration involves a point source of dimensional strength $m$, submerged a distance $L$ from the (undisturbed) surface, moving steadily through a fluid in a horizontal direction with speed $U$.  The motion of the source produces a steady ship wake pattern on the free surface.  We fix our frame of reference to move with the source, so that our governing equations are (\ref{eq:laplace})--(\ref{eq:farfield}), except that we do not have an additional pressure on the surface, so we set $\delta=0$ in (\ref{eq:dyncond}).  Further, we have the additional condition
	\begin{equation}
	\phi\sim x-\frac{\epsilon}{4\pi\sqrt{x^2+y^2+(z+1)^2}}\qquad\text{as }(x,y,z)\rightarrow(0,0,-1),\label{eq:sourcecond}
	\end{equation}
	where $\epsilon=m/(UL^2)$, which simply ensures that the velocity potential has the appropriate singular behaviour at the source.
	
	The linearised version of this problem ($\epsilon \ll 1$; see \S\ref{sec:linearsource}) is equivalent to flow past a submerged semi-infinite Rankine body with a rounded nose (like a cigar) which approaches a cylinder of radius $\sqrt{\epsilon/\pi}$ in the far-field \citep{batchelor67}.  Nonlinearity acts to distort the shape of the submerged body, and for highly nonlinear flows ($\epsilon \gg 1$), the analogy with a submerged Rankine body no longer holds \citep{pethiyagoda14b}.  In what follows, the numerical solutions of the fully nonlinear equations (\ref{eq:laplace})--(\ref{eq:farfield}) (with $\delta=0$) and (\ref{eq:sourcecond}) are computed using the boundary integral method outlined in \citet{pethiyagoda14a}, which is based on the algorithm outlined in \citet{forbes89}.
	
	\subsection{Flow past a submerged point doublet}
	\label{sec:flowdoublet}
	
	For our second flow configuration, we replace the point source singularity in \S\ref{sec:flowsource} with a point doublet of dimensional strength $\kappa$.  Thus, the governing equations are again (\ref{eq:laplace})--(\ref{eq:farfield}) with $\delta=0$, but instead of (\ref{eq:sourcecond}) we have
	\begin{equation}
	\phi\sim x+\frac{\mu x}{4\pi(x^2+y^2+(z+1)^2)^{3/2}},\qquad\text{as }(x,y,z)\rightarrow(0,0,-1),\label{eq:dipolecond}
	\end{equation}
	where $\mu=\kappa/(UL^3)$ is dimensionless strength of the doublet.
	
	In this case, the linear version of the problem ($\mu\ll 1$; see \S\ref{sec:lineardoublet}) is equivalent to flow past a submerged solid sphere of radius $(\mu/2\pi)^{1/3}$ \citep{lamb16}.  Nonlinear solutions for moderate values of $\mu$ have a very similar interpretation, except that the spherical body is distorted and indeed the surface is no longer closed \citep{pethiyagoda14b}.  All of our numerical solutions to the nonlinear problem (\ref{eq:laplace})--(\ref{eq:farfield}), (\ref{eq:dipolecond}) ($\delta=0$) are computed using the scheme outlined in \citet{pethiyagoda14b}.
	
	\subsection{Flow past a pressure distribution}
	\label{sec:flowpressure}
	
	Our third configuration involves studying the wake that forms behind a steadily moving pressure distribution applied to the free surface.  The type of pressure we are focussing on is localised and characterised by some pressure scale $P$ and horizontal length scale $L$.  By fixing our frame of reference to move with the pressure, the steady problem is to solve (\ref{eq:laplace})--(\ref{eq:farfield}), where $\delta=P/\rho U^2$ is the dimensionless pressure strength and $\rho$ is fluid density.
	
	In our formulation, the localised pressure distribution could take any form \citep{miao15}, provided it decays to zero as $x^2+y^2\rightarrow\infty$.  For the most part we use the simple Gaussian
	\begin{equation}
	p(x,y)=\mathrm{e}^{-\pi^2(x^2+y^2)},\label{eq:pressure}
	\end{equation}
	which has been used extensively in the past as a very simple model for generating ship wakes  \citep{darmon14,ellingsen14,li16,pethiyagoda15,pethiyagoda17,pethiyagoda18a,smeltzer17}.  In the present form, equations (\ref{eq:laplace})--(\ref{eq:farfield}), (\ref{eq:pressure}) together describe a nonlinear free-surface flow problem which depends on the Froude number $F$ and the pressure strength $\delta$.  In \S\ref{sec:slowlymovingobject}--\ref{sec:nonlinearflows} we present numerical results for this problem which are calculated using the numerical scheme outlined in \citet{pethiyagoda17}, which is based on earlier work by \citet{parau02,parau05a}.

	\subsection{Measuring the apparent wake angle}
	\label{sec:measuring}
	We consider two different methods of measuring the apparent wake angle of a ship wake, $\theta_\mathrm{app}$.  Method I was used by \citet{darmon14}, where a wave crest is isolated downstream and then the wave elevation is plotted against the polar angle as measured from the origin to the centreline ($y=0$). We choose an arbitrary percentage of the maximum height and mark the polar angle that gives this height as the apparent wake angle $\theta_\mathrm{app}$. For this paper we will use 5\%, 10\% and 20\% in our measurements ($\alpha=0.05$, $0.1$ and $0.2$).  We use Method I for low Froude numbers where the wake is dominated by transverse waves.
	
	The second method of measuring the apparent wake angle $\theta_\mathrm{app}$, Method II, given in \citet{pethiyagoda14b}, is performed by isolating each transverse wavelength of the wake and marking the highest peak. We then fit a line to the highest peaks using a linear least squares algorithm. The apparent wake angle $\theta_\mathrm{app}$ is then given by the angle between the fitted line and the centreline.  We use Method II for moderate values of the Froude number, for which wakes are made up of both transverse and divergent waves.
	
	\section{Linear regime}
	\label{sec:linear}
	
	\subsection{Submerged point source ($\epsilon\rightarrow 0$ with $F$ fixed)}\label{sec:linearsource}
	
	Taking the limit as $\epsilon\rightarrow 0$ while fixing $F$ has the effect of turning off the source while keeping its speed constant. We can linearise the problem about this limit to give a modified system of governing equations, namely Laplace's equation
	\begin{equation}
	\nabla^2\phi=0,\qquad\text{for }z<0,\label{eq:Linlaplace}
	\end{equation}
	together with the kinematic condition
	\begin{equation}
	\zeta_x=\phi_z \quad \text{on} \quad z=0,\label{eqn:NondimKinlinear}
	\end{equation}
	and the dynamic condition
	\begin{equation}
	\phi_x-1+\frac{\zeta}{F^2}=0 \quad\;\: \text{on} \quad z=0.\label{eqn:NondimDynlinear}
	\end{equation}
	The remaining governing equations (\ref{eq:radiation})--(\ref{eq:sourcecond}) remain unchanged.  Note that the kinematic and dynamic boundary conditions have been projected onto the plane $z=0$, which is a (well-known) key feature of this linearisation.
	
\begin{figure}
		\centering
		\subfloat[$F=0.1$]{\includegraphics[width=.32\linewidth]{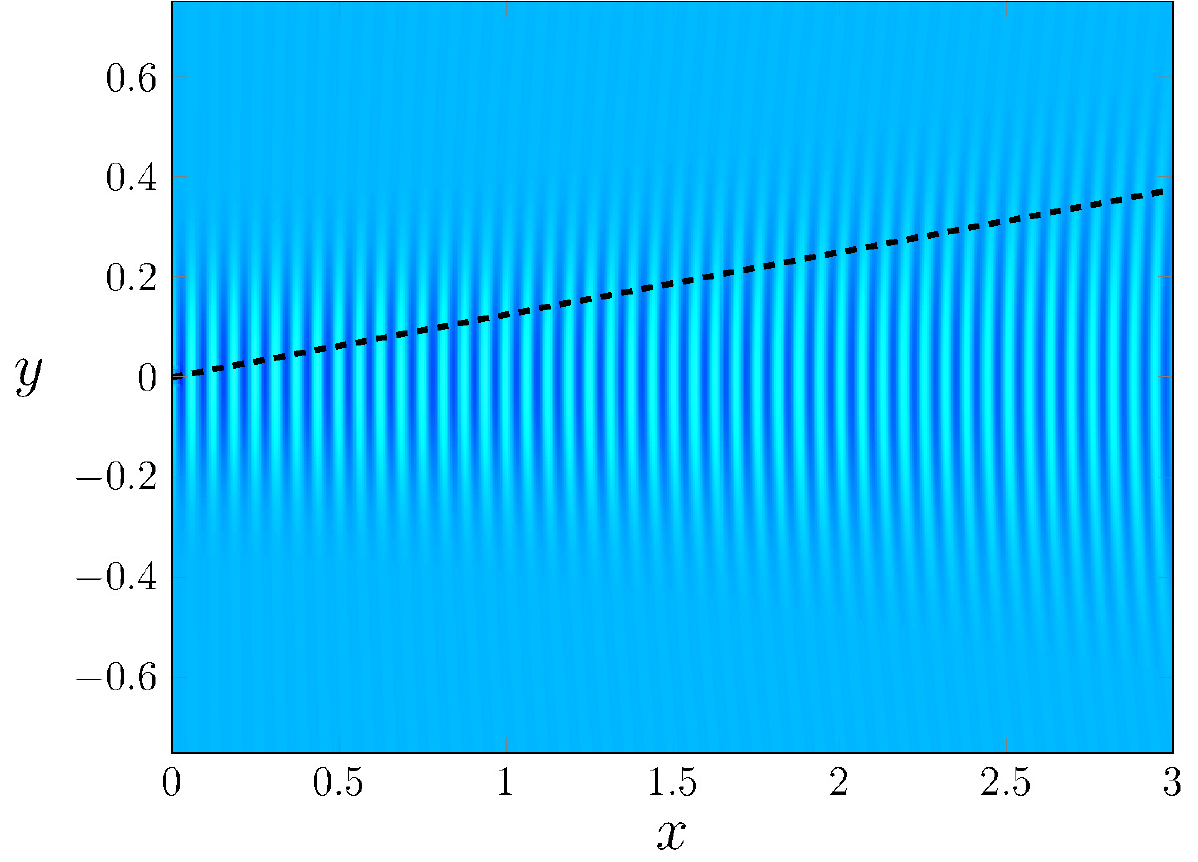}}\hspace{.01\linewidth}
		\subfloat[$F=0.2$]{\includegraphics[width=.32\linewidth]{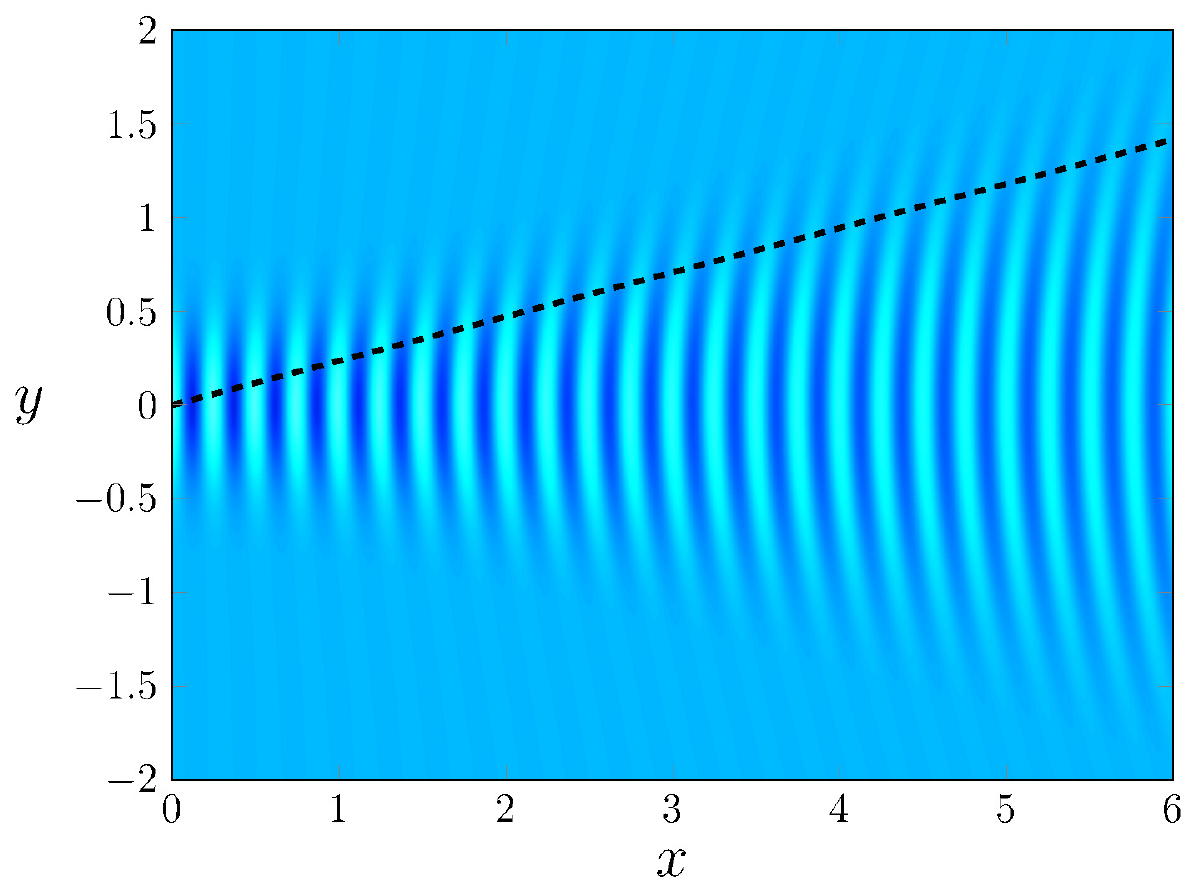}}\hspace{.01\linewidth}
		\subfloat[$F=1$]{\includegraphics[width=.325\linewidth]{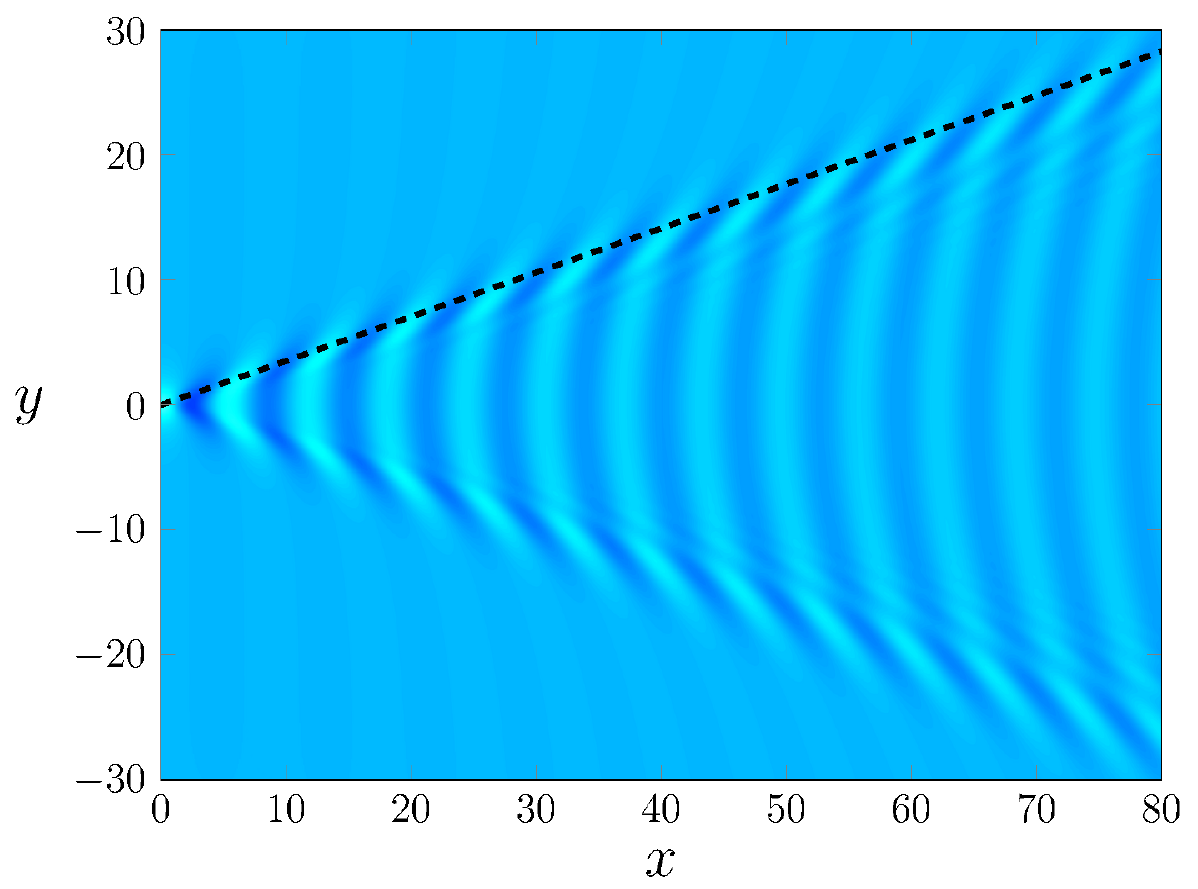}}\\
		\subfloat[$F=0.1$]{\includegraphics[width=.32\linewidth]{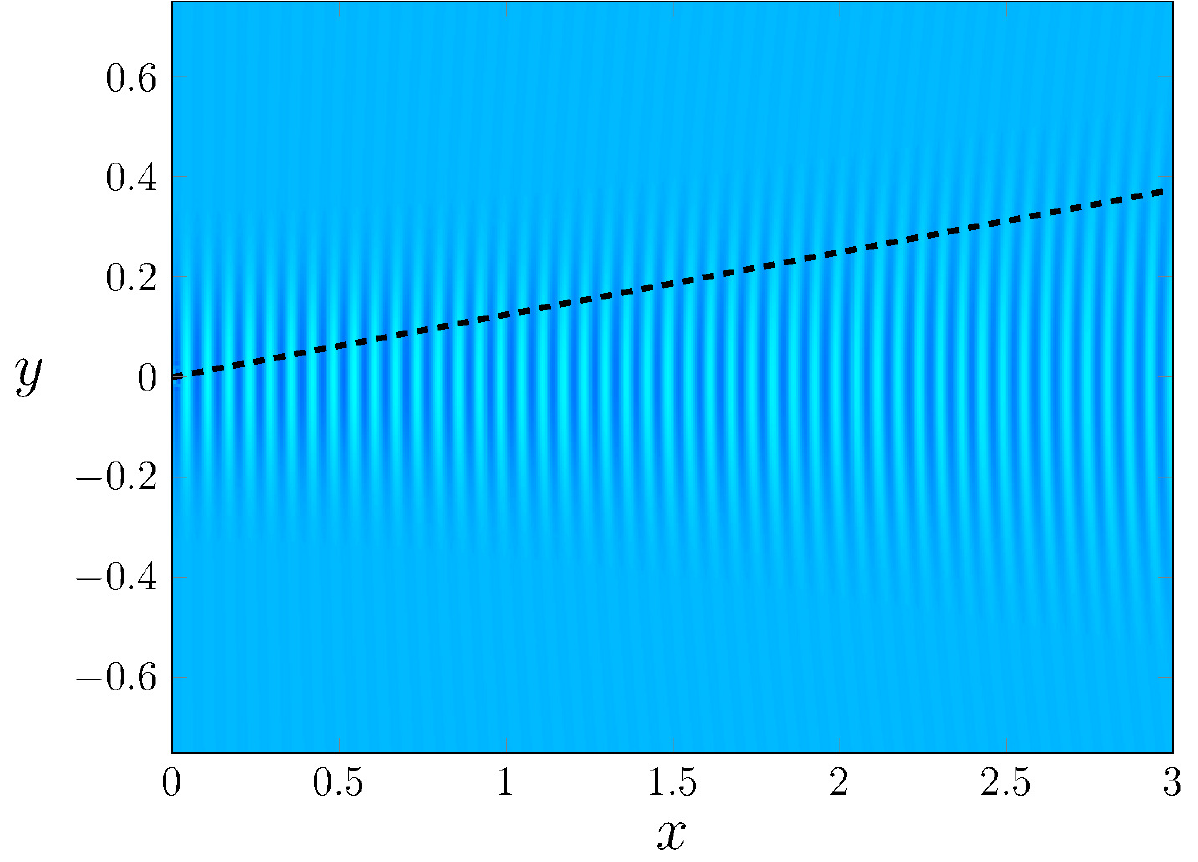}}\hspace{.01\linewidth}
		\subfloat[$F=0.2$]{\includegraphics[width=.32\linewidth]{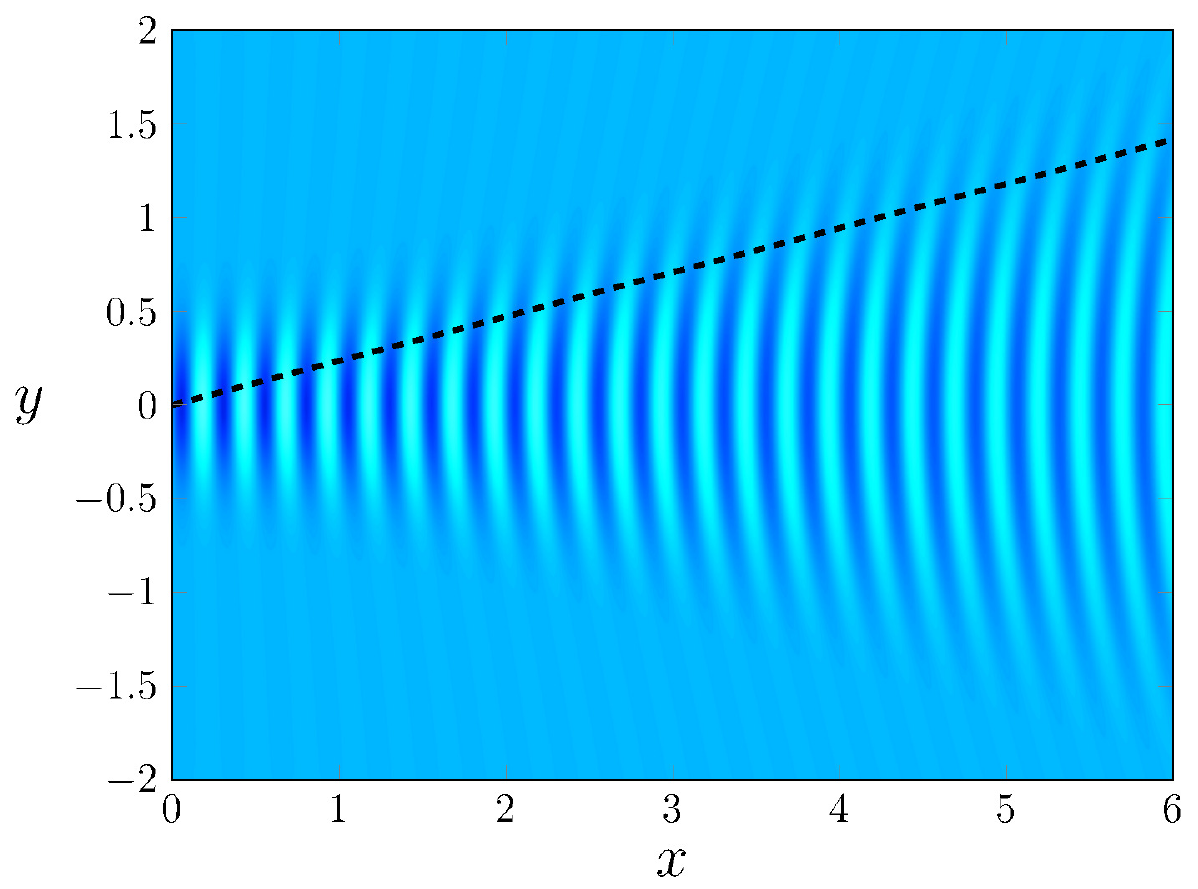}}\hspace{.01\linewidth}
		\subfloat[$F=1$]{\includegraphics[width=.325\linewidth]{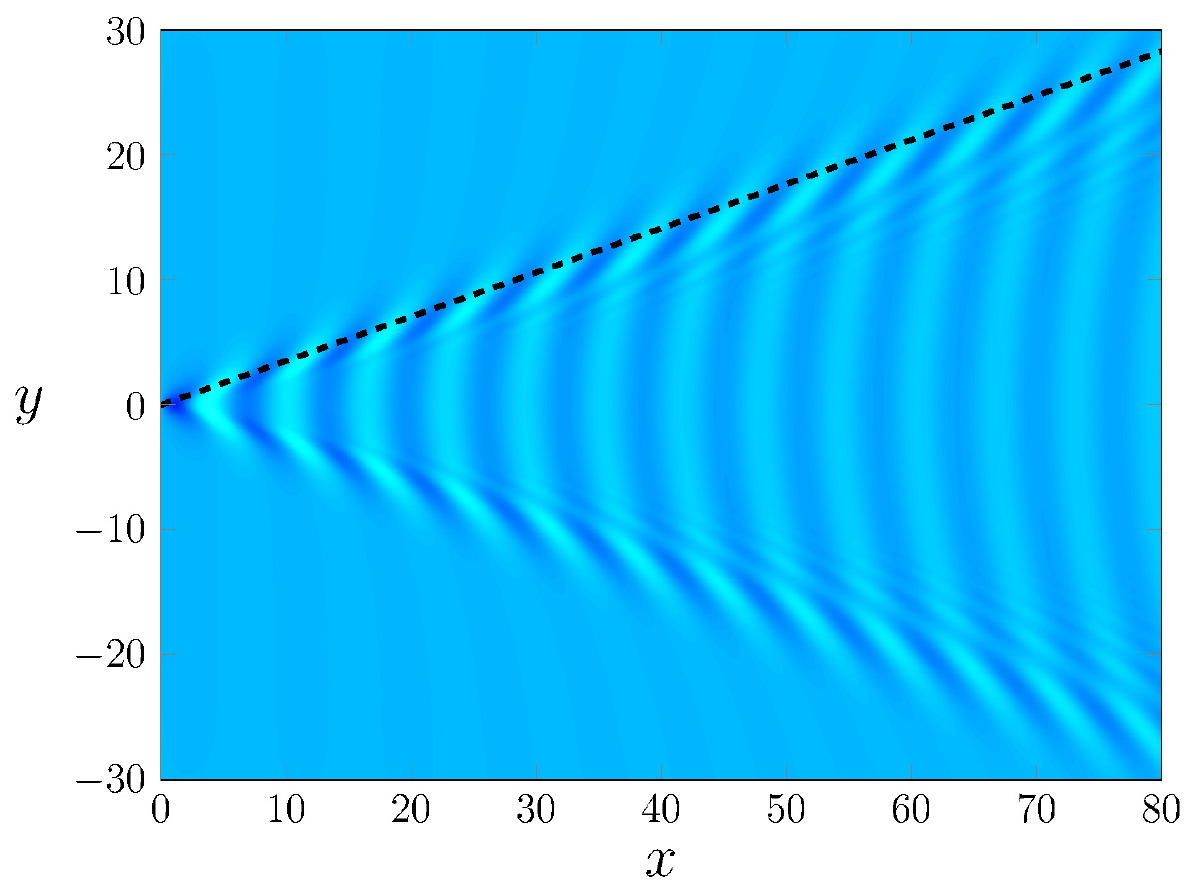}}\\
		\subfloat[Wave envelope]{\includegraphics[width=.8\linewidth]{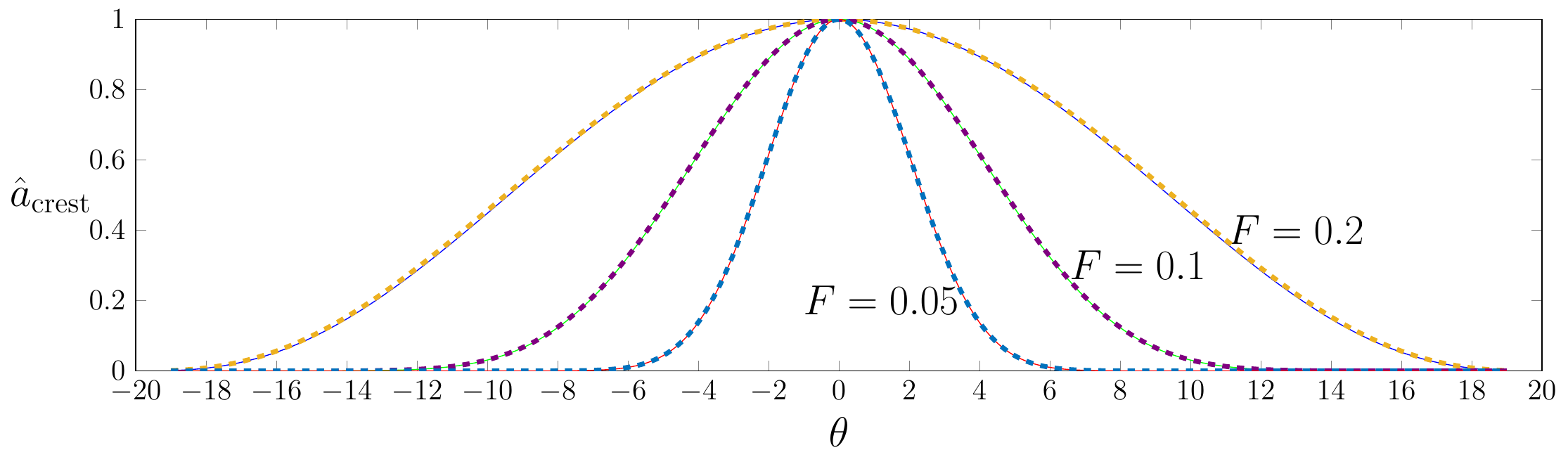}}
		\caption{(a)--(c) Plan view of the wave pattern for linear flow past a submerged source for Froude numbers $F=0.1$, $0.2$ and $1$.  Only the single integral in (\ref{eq:sourcefunc}) is used for the computation.  (d)-(f) Equivalent images for flow past a submerged doublet, this time computed using the single integral term of (\ref{eq:dipolefunc}).  In (a)-(b) and (d)-(e), the dashed line indicates the wake angle defined by 20\% of the wave height.  In (c) and (f), the dashed line is Kelvin's angle $\theta_{\mathrm{wedge}}$. (g) A plot of wave height against angle ($\theta$) for flow past a point source (solid curves) and a doublet (dashed curves) given by equations (\ref{eq:acrest}) and (\ref{eq:acrestdipole}), respectively, for the Froude numbers $F=0.2$ (blue, dashed orange), $F=0.1$ (green, dashed violet) and $F=0.05$ (red, dashed blue).}\label{fig:crestHight}
	\end{figure}

	We will use the exact solution to (\ref{eq:Linlaplace})-(\ref{eqn:NondimDynlinear}), (\ref{eq:radiation})--(\ref{eq:sourcecond}), given by \citet{peters49}, namely
	\begin{align}
	\zeta(x,y)=&-\frac{\epsilon F^2 \mathrm{sgn}(x)}{\pi^2}\int_{0}^{\frac{\pi}{2}}\cos\psi
	\int_{0}^{\infty}\frac{k \mathrm{e}^{-k|x|}\cos(k y \sin \psi)g(k,\psi)}{F^4 k^2+\cos^2\psi}\,\,\mbox{d}k\,\mbox{d}\psi
	\nonumber\\
	&+\frac{\epsilon H(x)}{\pi}\int_{-\infty}^{\infty}\xi \mathrm{e}^{-F^2\xi^2}\cos(x\xi)\cos(y\xi \lambda)\,\,\mbox{d}\lambda.
	\label{eq:sourcefunc}
	\end{align}
	where
	\begin{align}
	g(k,\psi)&=F^2 k \sin(k\cos \psi)+\cos\psi\cos(k\cos\psi).\label{eq:g}\\
	\xi(\lambda)&=\sqrt{\lambda^2+1}/F^2,\label{eq:xi}
	\end{align}
	{and $H(x)$ is the Heaviside function.} In this form, the double integral provides the near field component that decays rapidly away from the origin, while the single integral provides the wave train component.  In Figure~\ref{fig:crestHight}(a)-(c) we show the wave pattern from the single integral for three different Froude numbers.  For $F=0.1$ and $0.2$, we can see the wake is completely dominated by transverse waves, while for $F=1$ we see a mix of both transverse and divergent waves.  Note that the reason we only show the wave pattern in Figure~\ref{fig:crestHight} from the single integral is that, for small Froude number $F$, the near field contribution from the double integral is much larger than the wave amplitude, and so when we plot the full surface the waves are obscured by the near-field disturbance.

	
	
	
	We perform a stationary phase approximation on the exact linear solution (\ref{eq:sourcefunc}) which results in the approximate wave profile given in polar coordinates ($r=\sqrt{x^2+y^2}$, $\theta=\tan^{-1}(y/x)$):
	\begin{equation}
	\zeta(r,\theta) \sim a_1(r,\theta)\cos\left(r g(\lambda_1(\theta),\theta)+\frac{\pi}{4}\right)
	+a_2(r,\theta)\cos\left(r g(\lambda_2(\theta),\theta)-\frac{\pi}{4}\right)
	\label{eq:sphase}
	\end{equation}
	as $r\rightarrow\infty$ for $|\theta|<\arcsin(1/3)$, where
	\begin{align}
	\lambda_1(\theta)&=\frac{-1+\sqrt{1-8\tan^2\theta}}{4\tan\theta},
	&\lambda_2(\theta)&=\frac{-1-\sqrt{1-8\tan^2\theta}}{4\tan\theta},
	\label{eqn:parts1}\\
	f(\lambda)&=\frac{\epsilon\sqrt{\lambda^2+1}}{F^2}\,\mathrm{e}^{-\frac{\lambda^2+1}{F^2}},
	&g(\lambda,\theta)&=\frac{\sqrt{\lambda^2+1}}{F^2}
	(\cos\theta+\lambda\sin\theta),
	\label{eqn:parts2}\\
	a_1(r,\theta)&=\sqrt{\frac{2}{\pi}}\frac{f(\lambda_1(\theta))}{\sqrt{r|g_{\lambda\lambda}
			(\lambda_1(\theta),\theta)|}},
	&a_2(r,\theta)&=\sqrt{\frac{2}{\pi}}\frac{f(\lambda_2(\theta))}{\sqrt{r|g_{\lambda\lambda}
			(\lambda_2(\theta),\theta)|}},
	\label{eqn:parts3}
	\end{align}
	and $g_{\lambda\lambda}(\lambda,\theta)$ is the second partial derivative of $g(\lambda,\theta)$ with respect to $\lambda$.   In simple terms, the far-field surface takes the form
	\begin{equation}
	\zeta(r,\theta)\sim \frac{\epsilon}{r^{1/2}}{A}(\theta)\mathrm{e}^{-G(\theta)/F^2}
	\quad\mbox{as}\quad r\rightarrow\infty,
	\label{eq:roughfarfield}
	\end{equation}
	where ${A}$ and $G$ are functions of $\theta$ {\citep{pethiyagoda14b}}.  The linear dependence on $\epsilon$ in (\ref{eq:roughfarfield}) arises because we are considering the linear version of the problem.  The $r^{-1/2}$ dependence implies that the wave pattern decays rather slowly as we move further away from the cause of the waves.  Finally, the exponential dependence on $-F^{-2}$ reveals that the waves themselves become extremely small in the small Froude number limit, which is the limit we are most interested in here.
	
	We now consider (\ref{eq:sphase}) in more detail.  The factors $a_1$ and $a_2$ represent the function envelopes for the transverse and divergent waves, respectively.  Since $\lambda_1^2<\lambda_2^2$ for $0\leq \theta < \theta_{\mathrm{wedge}}$ (where remember that $\theta_{\mathrm{wedge}}$ is Kelvin's angle), it follows that $\mathrm{e}^{-(\lambda_2^2+1)/{F^2}}\ll \mathrm{e}^{-(\lambda_1^2+1)/{F^2}}$ for $F\ll 1$.  In other words, for small Froude numbers the transverse waves exponentially dominate the divergent waves and thus, in practice, we will observe the highest parts of the wave pattern along the crestlines of the transverse waves.
	
	The location of the crestlines is given by
	\begin{equation}
	\cos\left(r g(\lambda_1(\theta),\theta)+\frac{\pi}{4}\right)=1,
	\end{equation}
	which leads to the relationship
	\begin{equation}
	r(\theta)=\frac{(2n-1/4)\pi}{g(\lambda_1(\theta),\theta)},\label{eq:crestPos}
	\end{equation}
	where $n$ is some positive integer. Thus, we have a formula for the wave height along the crest:
	\begin{equation}
	a^s_\mathrm{crest}(\theta)=\epsilon\sqrt{\frac{2g(\lambda_1(\theta),\theta)}{(2n-1/4)\pi^2g_{\lambda\lambda}
			(\lambda_1(\theta),\theta)}}\,f(\lambda_1(\theta)).
	\end{equation}
	Scaling such that $\hat{a}^s_\mathrm{crest}(0)=1$ gives:
	\begin{equation}
	\hat{a}^s_\mathrm{crest}(\theta)=(\lambda_1(\theta)^2+1)^\frac{3}{2}\sqrt{\frac{\cos\theta+\lambda_1(\theta)\sin\theta}{\cos\theta+(2\lambda_1(\theta)^3+3\lambda_1(\theta))
			\sin\theta}}\,\mathrm{e}^{-{\lambda_1(\theta)^2}/{F^2}}.\label{eq:acrest}
	\end{equation}
	Plotting this crest height for different Froude numbers (Figure \ref{fig:crestHight}(g)), we see that as the Froude number $F$ decreases, the Gaussian-like wave crest narrows, which means that the apparent wake angle decreases as $F$ decreases. We can use (\ref{eq:acrest}) to determine the wake angle, $\theta_\mathrm{app}$, by setting $\hat{a}^s_\mathrm{crest}(\theta_\mathrm{app})=\alpha$, where $\alpha$ is the chosen fraction of the maximum wave height.  For example, in Figure \ref{fig:crestHight}(a)-(b) we have plotted dashed lines that indicate the apparent wake angle when $\alpha$ is chosen to be $\alpha=0.2$.


	\subsection{Submerged point doublet ($\mu\rightarrow 0$ with $F$ fixed)}\label{sec:lineardoublet}

	The linear governing equations for flow past a submerged doublet are given by (\ref{eq:Linlaplace})--(\ref{eqn:NondimDynlinear}), (\ref{eq:radiation})--(\ref{eq:farfield}) and (\ref{eq:dipolecond}).  They are derived by fixing $F$ in the fully nonlinear equations and taking the limit $\mu\rightarrow 0$, which is equivalent to considering flow past a submerged sphere of vanishing radius while keeping its depth and the far-field speed constant.  The exact solution to the linear problem is given by
	\begin{align}
	\zeta(x,y)=&\frac{\mu F^2}{\pi^2}\int_{0}^{\frac{\pi}{2}}\cos\psi
	\int_{0}^{\infty}\frac{k^2 \mathrm{e}^{-k|x|}\cos(k y \sin \psi)g(k,\psi)}{F^4 k^2+\cos^2\psi}\,\,\mbox{d}k\,\mbox{d}\psi
	\nonumber\\
	&-\frac{\mu H(x)}{\pi}\int_{-\infty}^{\infty}\xi^2 \mathrm{e}^{-F^2\xi^2}\sin(x\xi)\cos(y\xi \lambda)\,\,\mbox{d}\lambda,
	\label{eq:dipolefunc}
	\end{align}
	where $g(k,\psi)$ and $\xi(\lambda)$ are given by (\ref{eq:g}) and (\ref{eq:xi}), respectively.  Representative wave patterns computed using the single integral term in (\ref{eq:dipolefunc}) are shown in Figure~\ref{fig:crestHight}(d)-(f).  For the two small Froude numbers $F=0.1$ and $0.2$, the plan view looks very similar to that for flow past a submerged source, with the wake appearing to be made up of transverse waves only, and the waves themselves appearing to be a phase-shift when compared to flow past a submerged source.  For the moderate Froude number $F=1$, the image is more like a traditional ship wake, with both transverse and divergent waves.
	
	As with the flow due to a submerged source, we can write the stationary phase approximation
	\begin{equation}
	\zeta(r,\theta) \sim a_1(r,\theta)\sin\left(r g(\lambda_1(\theta),\theta)+\frac{\pi}{4}\right)
	+a_2(r,\theta)\sin\left(r g(\lambda_2(\theta),\theta)-\frac{\pi}{4}\right),
	\label{eq:dipolesphase}
	\end{equation}
	where $a_1(r,\theta)$, $a_2(r,\theta)$, $\lambda_1(\theta)$, $\lambda_2(\theta)$ and $g(\lambda,\theta)$ are given by (\ref{eqn:parts1})--(\ref{eqn:parts3}), but this time
	\begin{equation}
	f(\lambda)=\mu(\lambda^2+1)\mathrm{e}^{-{(\lambda^2+1)}/{F^2}}.\label{eqn:parts4}
	\end{equation}
	Following the same argument, the scaled amplitude along the transverse wave crest is given by
	\begin{equation}
	\hat{a}^d_\mathrm{crest}(\theta)=(\lambda_1(\theta)^2+1)^2\sqrt{\frac{\cos\theta+\lambda_1(\theta)\sin\theta}{\cos\theta+(2\lambda_1(\theta)^3+3\lambda_1(\theta))
			\sin\theta}}
	\,\mathrm{e}^{-{\lambda_1(\theta)^2}/{F^2}}.\label{eq:acrestdipole}
	\end{equation}
	We note that $\hat{a}^d_\mathrm{crest}(\theta)=\sqrt{(\lambda_1(\theta)^2+1)}\hat{a}^s_\mathrm{crest}(\theta)$, and $\lambda_1\rightarrow 0$ as $F\rightarrow 0$; therefore, the shape of the transverse wave crest for flow due to a submerged doublet will approach that due to a submerged source in the  low Froude number limit. This property can be seen in Figure~\ref{fig:crestHight}(g), where the curves are actually different but virtually indistinguishable on this scale.
	
	
	\subsection{Applied pressure distribution ($\delta\rightarrow 0$ with $F$ fixed)}
	
	The linear governing equations for flow past a pressure distribution are given by (\ref{eq:Linlaplace})--(\ref{eqn:NondimKinlinear}), (\ref{eq:radiation})--(\ref{eq:farfield}) and the linearised dynamic condition
	\begin{equation}
	\phi_x-1+\frac{\zeta}{F^2}+p(x,y)=0 \quad\;\: \text{on} \quad z=0,\label{eqn:NondimDynlinearperss}
	\end{equation}
	where the applied pressure distribution is (\ref{eq:pressure}).  One form for the exact solution{, adapted from \citet{pethiyagoda17},} is
	\begin{align}
	\zeta(x,y) = &-\delta F^2 p(x,y)+\frac{\delta F^2}{2\pi^3} \int\limits_{-\pi/2}^{\pi/2}\,\int\limits_{0}^{\infty}\frac{k^2\mathrm{e}^{-k^2/4\pi^2}\cos(k[|x|\cos\psi+y\sin\psi])}{k-k_0}\,\,\mathrm{d}k\,\,\mathrm{d}\psi\notag\\
	&-\frac{\delta F^2 H(x)}{\pi^2}\int_{-\infty}^{\infty}\xi^2 \mathrm{e}^{-F^4\xi^4/4\pi^2}\sin(x\xi)\cos(y\xi \lambda)\,\,\mbox{d}\lambda,\label{eq:exactPressureLinearInfinite}
	\end{align}
	where the path of $k$-integration is diverted below the pole $k=1/(F^2\cos^2\psi)$.
	
	As with the submerged doublet, the stationary phase approximation is given by (\ref{eq:dipolesphase}), where $a_1(r,\theta)$, $a_2(r,\theta)$, $\lambda_1(\theta)$, $\lambda_2(\theta)$ and $g(\lambda,\theta)$ are given by (\ref{eqn:parts1})--(\ref{eqn:parts3}),
	\begin{equation}
	f(\lambda)=\delta F^2(\lambda^2+1)\mathrm{e}^{-{(\lambda^2+1)^2}/{4\pi^2F^4}}\label{eqn:pressureF}
	\end{equation}
	and the scaled transverse wave crest is given by
	\begin{equation}
	\hat{a}^p_\mathrm{crest}(\theta)
	=(\lambda_1(\theta)^2+1)^2\sqrt{\frac{\cos\theta+\lambda_1(\theta)\sin\theta}{\cos\theta+(2\lambda_1(\theta)^3+3\lambda_1(\theta))\sin\theta}}
	\,\mathrm{e}^{-{((\lambda_1(\theta)^2+1)^2-1)}/{4\pi^2F^4}}.\label{eq:acrestpressure}
	\end{equation}
	
	As mentioned above, we are using the applied pressure distribution (\ref{eq:pressure}) to align this work with a number of previous studies.  It is interesting to note that the choice
	\begin{equation}	
	p(x,y)=-\frac{\partial}{\partial x}\left(\frac{1}{2\pi\sqrt{x^2+y^2+1}}\right),
    \label{eq:pressuresource}
	\end{equation}
	provides a wave pattern that is identical to that produced by a submerged source. Similarly,
	\begin{equation}
	p(x,y)=-\frac{\partial^2}{\partial x^2}\left(\frac{1}{2\pi\sqrt{x^2+y^2+1}}\right),
    \label{eq:pressuredoublet}
	\end{equation}
	provides a wave pattern that is identical to that produced by a submerged doublet \citep{mccue19}, {where one of the two terms after performing differentiation was shown by \citet{havelock19} to have the same far field behaviour as flow past a doublet.}
	
	\subsection{Apparent wake angle for linear flows}
	
\begin{figure}
		\centering
		\includegraphics[width=.65\linewidth]{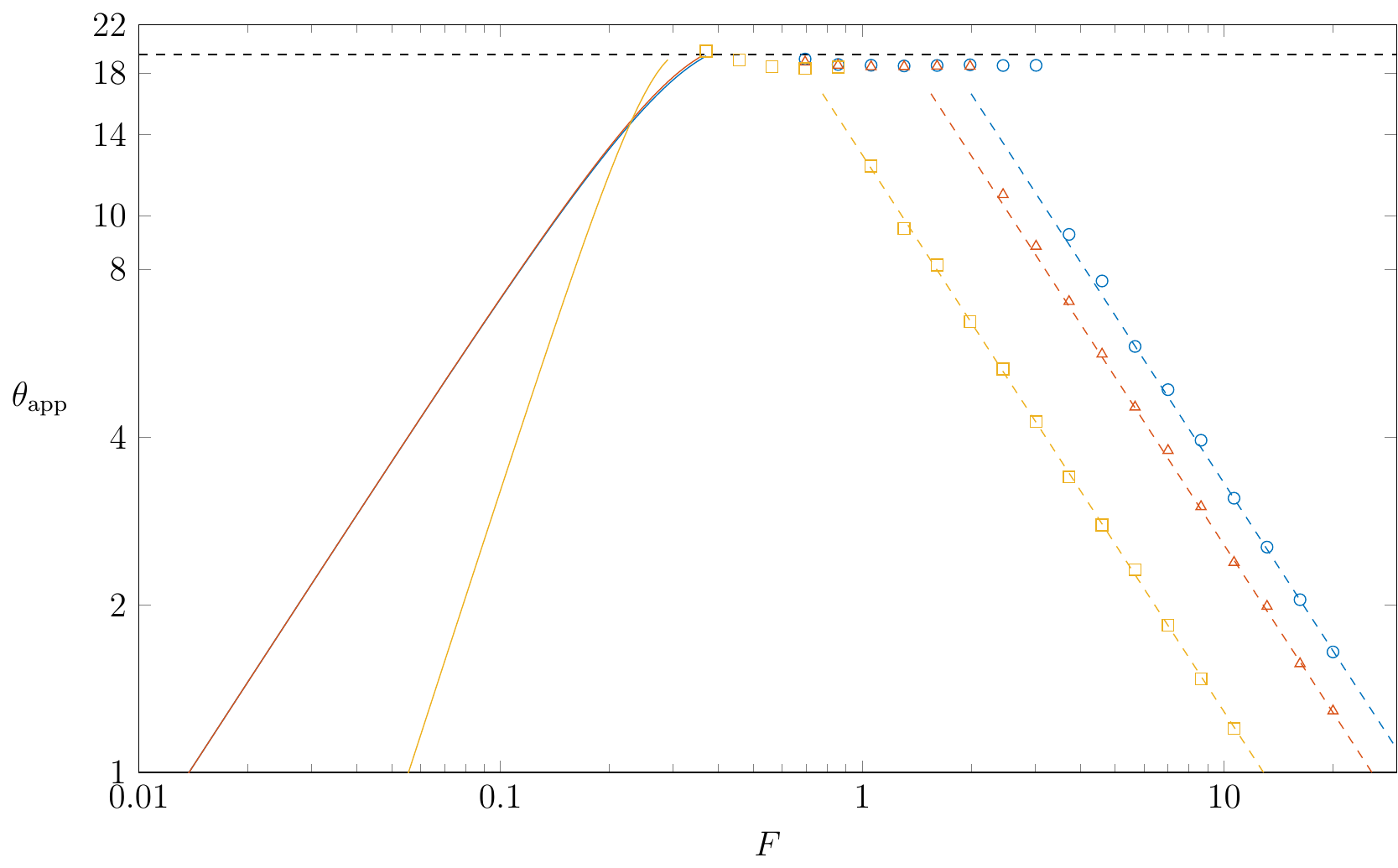}
		\caption{Plots of apparent wake angle against the Froude number. The solid curve represents method I defined in \S\ref{sec:measuring} using a 20\% cutoff height. The low Froude number results only consider the transverse waves of the stationary phase approximation. The apparent wake angles calculated from method II are given by the marks with the large Froude approximation given by the dashed line. The blue, red and yellow colours denote flow past a submerged source, submerged doublet and applied pressure distribution, respectively.}\label{fig:wakeAngle}
	\end{figure}

	In Figure \ref{fig:wakeAngle} we plot apparent wake angles for linearised flow past a submerged source (blue), submerged doublet (red) or an applied pressure distribution (violet).  The results using Method I (see \S\ref{sec:measuring}), valid for low Froude numbers, are represented by the solid curves for a 20\% cutoff height ($\alpha=0.2$).  The open circles, triangles and squares are computed using Method II.  The dashed lines show the large Froude number approximation for Method II found using the method of stationary phase.
	
	We see from Figure \ref{fig:wakeAngle} that for low Froude numbers the apparent wake angle $\theta_\mathrm{app}$ appears to depend linearly on the Froude number $F$ on this log scale, which suggests a power-law relationship.  For flow past a point source we can determine the relationship by first setting the desired amplitude $\hat{a}^s_\mathrm{crest}=\alpha$, taking logs of both sides and assuming $\theta_\mathrm{app}\ll 1$ to give
	\begin{equation}
	\ln\alpha=\left(\frac{5}{2}-\frac{1}{F^2}\right)\theta^2+\mathcal{O}(\theta^4).\label{eq:lowFroudeAsymptStep}
	\end{equation}
	Rearranging (\ref{eq:lowFroudeAsymptStep}) for $\theta_\mathrm{app}$ and taking $F\rightarrow 0$, we find
	\begin{equation}
	\theta_\mathrm{app}\sim \sqrt{\ln(1/\alpha)}\,F\quad\text{as }F\rightarrow 0. \label{eq:lowFroudeAsymptsd}
	\end{equation}
	Performing the same procedure to flow past a doublet will give the same scaling (\ref{eq:lowFroudeAsymptsd}), while for flow past the pressure distribution (\ref{eq:pressure}) we find
	\begin{equation}
	\theta_\mathrm{app}\sim \pi\sqrt{2\ln(1/\alpha)}\,F^2\quad\text{as }F\rightarrow 0.
	\label{eq:lowFroudeAsymptpressure}
	\end{equation}
	{ We note that while the wake angle scaling for flow past the pressure distribution (\ref{eq:pressure}) is different from the scaling for flow past a source/doublet, different pressures can lead to different scalings. In fact we have reported pressure distributions in \S\ref{sec:slowpressure} that will give the same wake pattern and thus the same wake angle scaling as flow past a submerged source/doublet. Additionally, in Appendix \ref{sec:powerLawPress} we show that the wake angle scaling for a general pressure distribution depends heavily on the Fourier transform and its derivatives.} In a sense, Figure \ref{fig:wakeAngle} combines the results of \cite{darmon14} and \cite{pethiyagoda14b} (which are for moderate to high Froude numbers) and fills in the gaps by including our new results for smaller Froude numbers.
	
	
	
	\subsection{Wake envelope via exponential asymptotics}
	
	In our analysis above we have applied the method of stationary phase to determine the far-field wake and, subsequently, the apparent wake angle $\theta_\mathrm{app}$, for linear flow past a submerged point source, submerged doublet and applied pressure distribution.  In this subsection we concentrate on flow past a submerged source or doublet, and apply ideas from exponential asymptotics to derive an envelope function which encloses the low-Froude number wake in both the near and far field.
	
	We suspect from our stationary phase argument that the wave amplitude for the linear problems is of order $\mathrm{e}^{-\chi/F^2}$ for $F\ll 1$, where $\chi$ is a function of position.  A straightforward asymptotic expansion implies  that, for small $F$, the surface height can be represented by
	\begin{equation}
	\zeta(x,y)=\sum_{n=0}^{N-1} F^{2n} \zeta_n(x,y) + Z_N(x,y),
	\label{eq:zetaseries}
	\end{equation}
	where $Z_N(x,y)$ is the remainder after $N$ terms.  If the summation in (\ref{eq:zetaseries}) is taken to infinity, the series is divergent; however, if $N$ is chosen carefully so that the series is truncated optimally (at the least term) \citep{boyd99}, then
	\begin{equation}
	Z_N \sim {B}(x,y) \mathrm{e}^{-\chi/F^2}\quad\mbox{as}\quad F\rightarrow 0.\label{eq:Remainder}
	\end{equation}
	In other words, in order to capture the wave pattern whose amplitude is smaller than all orders of the traditional asymptotic series in powers of $F^2$, we must analyse the remainder term after the original series is truncated optimally.  A leading order expression for $Z_N$ is called a superasymptotic approximation~\citep{berry91}.
	
	The key issue here is that the singulant function $\chi$ governs how small $Z_N$ will be.  In particular, as $\mathrm{Re}(\chi)$ increases, then $Z_N$ decreases exponentially.  Thus, we are motivated to plot curves of constant $\mathrm{Re}(\chi)$, with the goal of deriving an effective envelope (or boundary) of the wake for small Froude number.  For the geometry of flow past a submerged point source, this problem has been studied by \citet{lustri13}.
{We will not repeat the details here as they are extensive, except to say that in order to derive a PDE for $\chi$, they write out an ansatz for the terms $\zeta_n$ in the limit $n\rightarrow$ and (together with an analogous expression for the late order terms in a series for $\phi$) substitute this into the governing equations to a coupled system of PDEs.  These in turn combine to give}
	\begin{equation}
    \left(\frac{\partial\chi}{\partial x}\right)^4+\left(\frac{\partial\chi}{\partial x}\right)^2+\left(\frac{\partial\chi}{\partial y}\right)^2=0
    \label{eq:SingulantEq}
	\end{equation}
	subject to the condition that $\chi$ must vanish at the singularity of the analytically-continued free surface ($x,y \in \mathbb{C}$, $z =0$), namely
	\begin{equation}
	\chi = 0 \qquad \mathrm{on} \qquad x^2 + y^2 + 1 = 0.\label{eq:SingulantBC}
	\end{equation}
	{Using Charpit's method, this PDE can be solved to give}
	\begin{equation}
	\chi=\pm\frac{s-x}{s(2+s^2)},\label{eq:u}
	\end{equation}
	where $s$ is one of the four roots to the quartic
	\begin{equation}
	(x^2+y^2)s^4+4x s^3+(x^2+4y^2+4)s^2+4x s+(4y^2+4)=0.\label{eq:sroot}
	\end{equation}
	We are interested in the curves defined by
	\begin{equation}
	\mathrm{Re}(\chi)=\beta \label{eq:uCurve}
	\end{equation}
	where $\beta$ is some constant, the positive value of $\chi$ and $s$ such that $\mathrm{Re}(x s)>0$ is chosen. There is no need to distinguish between complex conjugates.

	\begin{figure}
		\centering
		\subfloat[$F=0.1$]{\includegraphics[width=.45\linewidth]{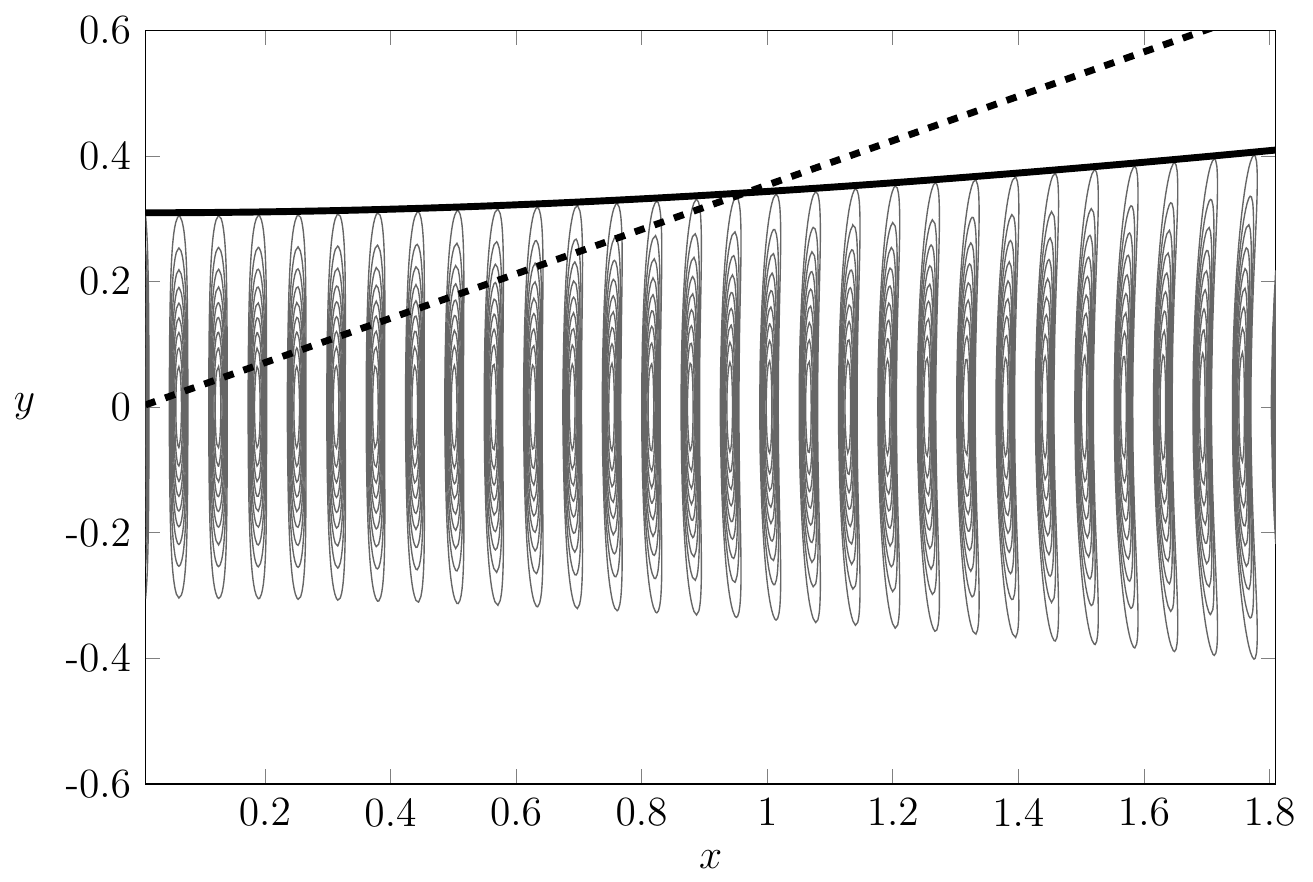}}\hspace{.05\linewidth}
		\subfloat[$F=0.2$]{\includegraphics[width=.45\linewidth]{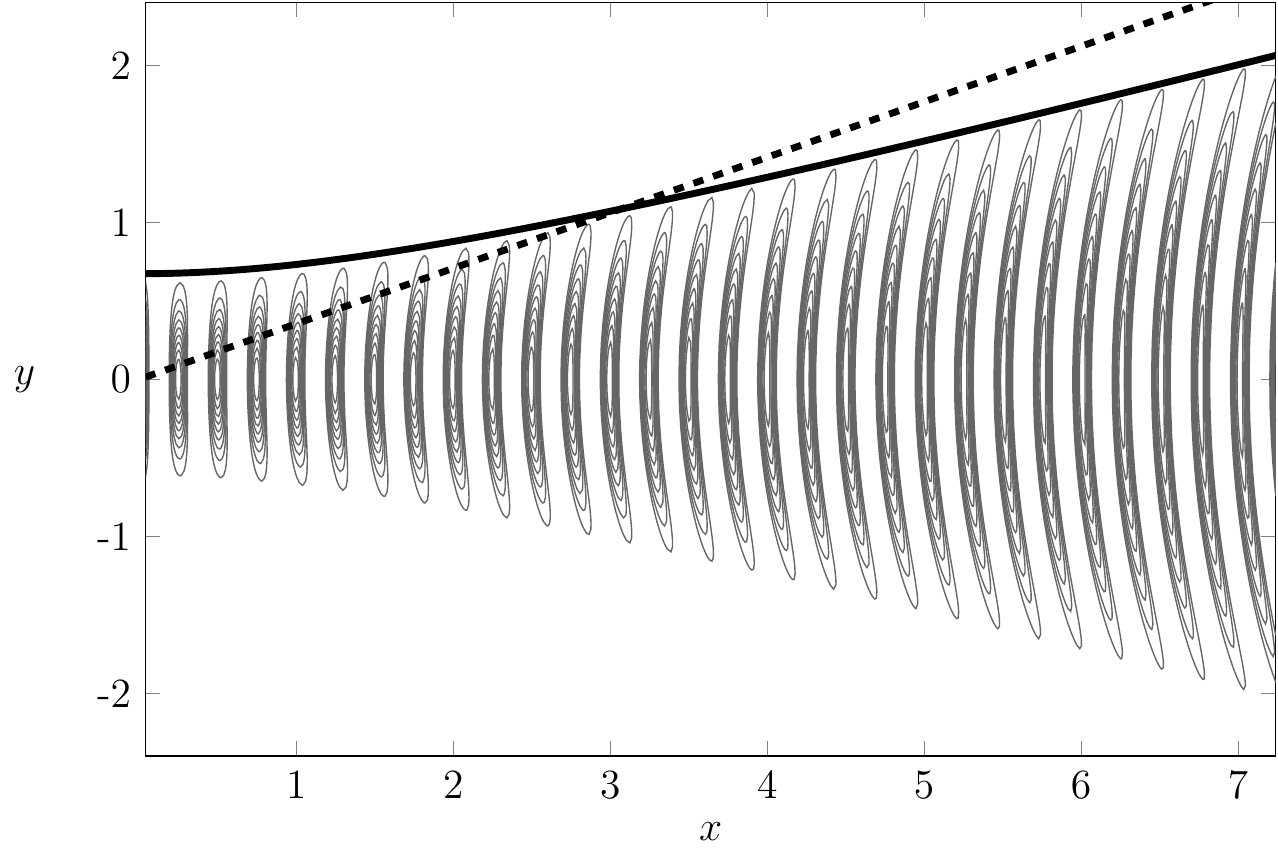}}\\
		\subfloat[$F=0.4$]{\includegraphics[width=.45\linewidth]{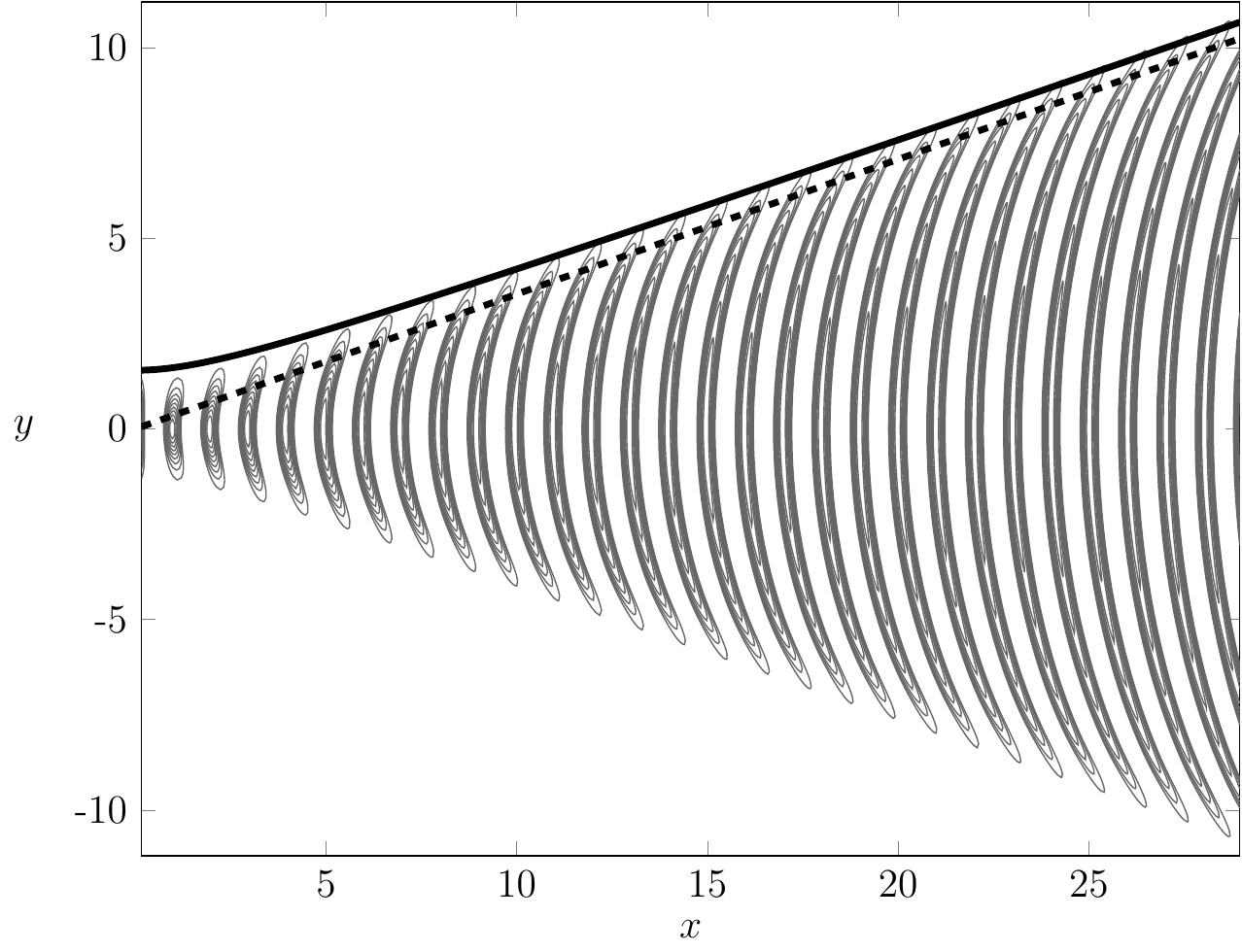}}\hspace{.05\linewidth}
		\subfloat[$F=0.8$]{\includegraphics[width=.45\linewidth]{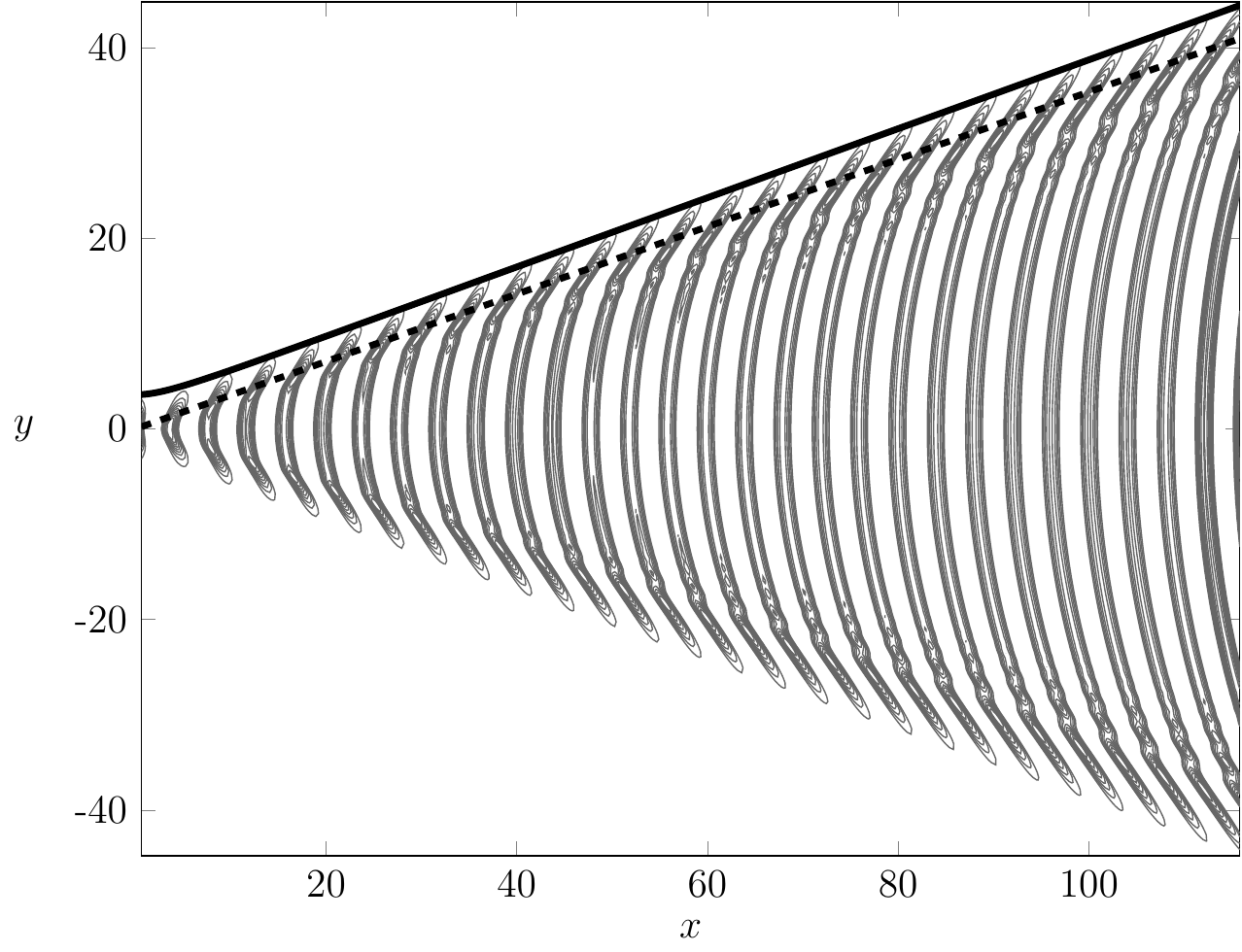}}
		\caption{A contour plot of the positive heights of the free surface profile for flow past a submerged source, in which there is a contour for every 10th percentile of the wave height within a single wavelength. The solid line is a line of constant $\mathrm{Re}(\chi)$ as given in (\ref{eq:u}), specifically (a) $\mathrm{Re}(\chi)= 1.0236$ (b) $\mathrm{Re}(\chi)= 1.1026$ (c) $\mathrm{Re}(\chi)= 1.4170$ and (d) $\mathrm{Re}(\chi)= 2.3602$. The dashed line is Kelvin's angle $\theta_{\mathrm{wedge}}$ measured from the origin.}\label{fig:Contour}
	\end{figure}
	
	To relate (\ref{eq:uCurve}) to the apparent wake angle $\theta_\mathrm{app}$ we consider the leading order asymptotic approximation {of $\mathrm{Re}(\chi)$} for $r\rightarrow\infty$ (in polar coordinates), {derived in Appendix \ref{sec:chiApprox}},
	\begin{equation}
	\mathrm{Re}(\chi)=f(\theta)+\mathcal{O}(r^{-1}),\label{eq:uExpand}
	\end{equation}
	where
	\begin{equation} f=\frac{3\cos\theta\sqrt{9\cos^2\theta-8}-9\cos^2\theta+8}
	{27\cos^4\theta-9\cos^3\theta\sqrt{9\cos^2\theta-8}-42\cos^2\theta+10\cos\theta\sqrt{9\cos^2\theta-8}+16}.\label{eq:uApprox}
	\end{equation}
	Therefore, for some constant $\beta$, the curve $\mathrm{Re}(\chi)=\beta$ approaches a ray of angle $\theta$ measured from the origin defined as the solution to $f(\theta)=\beta$.  In Figure~\ref{fig:Contour} we present contour plots, where the contours are given for every 10th percentile of the wave height within a single wavelength, for the free-surface profiles given by the single integral term of (\ref{eq:sourcefunc}).  We also include the curve defined by $\mathrm{Re}(\chi)=\beta$ where for (a)-(b) $f(\theta_\mathrm{app})=\beta$, $\hat{a}_\mathrm{crest}(\theta_\mathrm{app})=\alpha$ and $\alpha=0.1$. For Figure~\ref{fig:Contour}(c)-(d) the constant $\beta$ is chosen by ensuring the curve passes through the edge of the 10\% ($\alpha=0.1$) contour. We can see that the curve (\ref{eq:uCurve}) closely borders the tenth percentile contour.  {Note that we have drawn plots like those in Figure~\ref{fig:Contour} using different values of $\alpha$ (not shown here) and the agreement is just as good as that shown in Figure~\ref{fig:Contour}.}
	
\begin{figure}
		\centering
		\includegraphics[width=.8\linewidth]{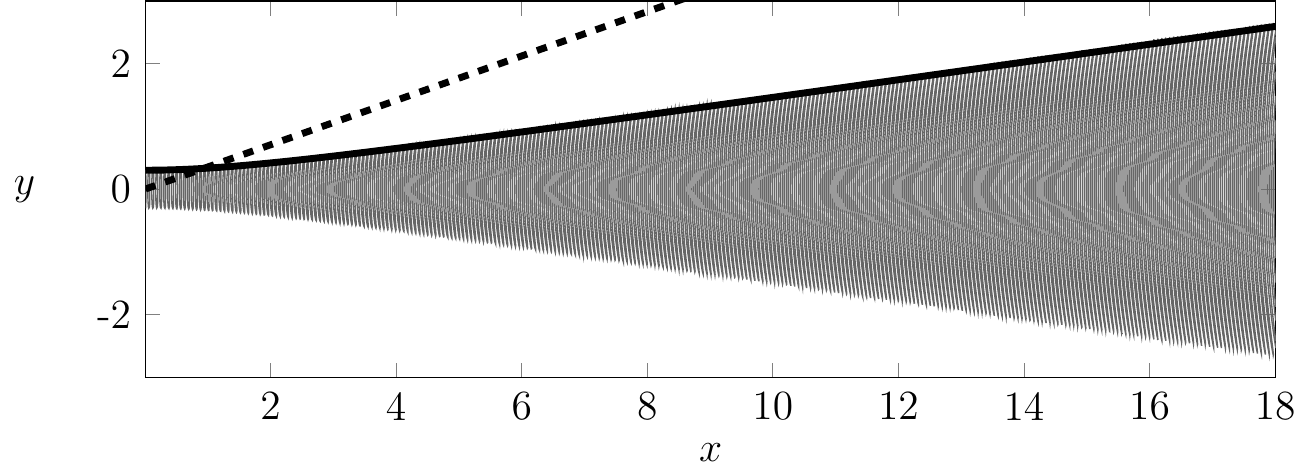}
		\caption{A contour plot of the positive heights of the free surface profile for flow past a source with $F=0.1$, The contour is at 10\% of the wave height within a single wavelength. The solid line is given by $\mathrm{Re}(\chi)= 1.0236$. The dashed line is Kelvin's angle $\theta_{\mathrm{wedge}}$.}\label{fig:ContourLarge}
	\end{figure}

	The image in Figure~\ref{fig:Contour}(a), which is for a very small Froude number $F=0.1$, focuses on the near field wave pattern.  To appreciate the sense in which the curve $\mathrm{Re}(\chi)=\beta$ bends around and approaches a constant ray in the far-field, we show in Figure~\ref{fig:ContourLarge} the same image but drawn on a much larger scale.  Taking together, both Figure~\ref{fig:Contour}(a) and Figure~\ref{fig:ContourLarge} demonstrate how well the curve $\mathrm{Re}(\chi)=\beta$ acts to define the envelope of the wave pattern for small Froude numbers.
	
\begin{figure}
		\centering
		\subfloat[$F=0.1$]{\includegraphics[width=.45\linewidth]{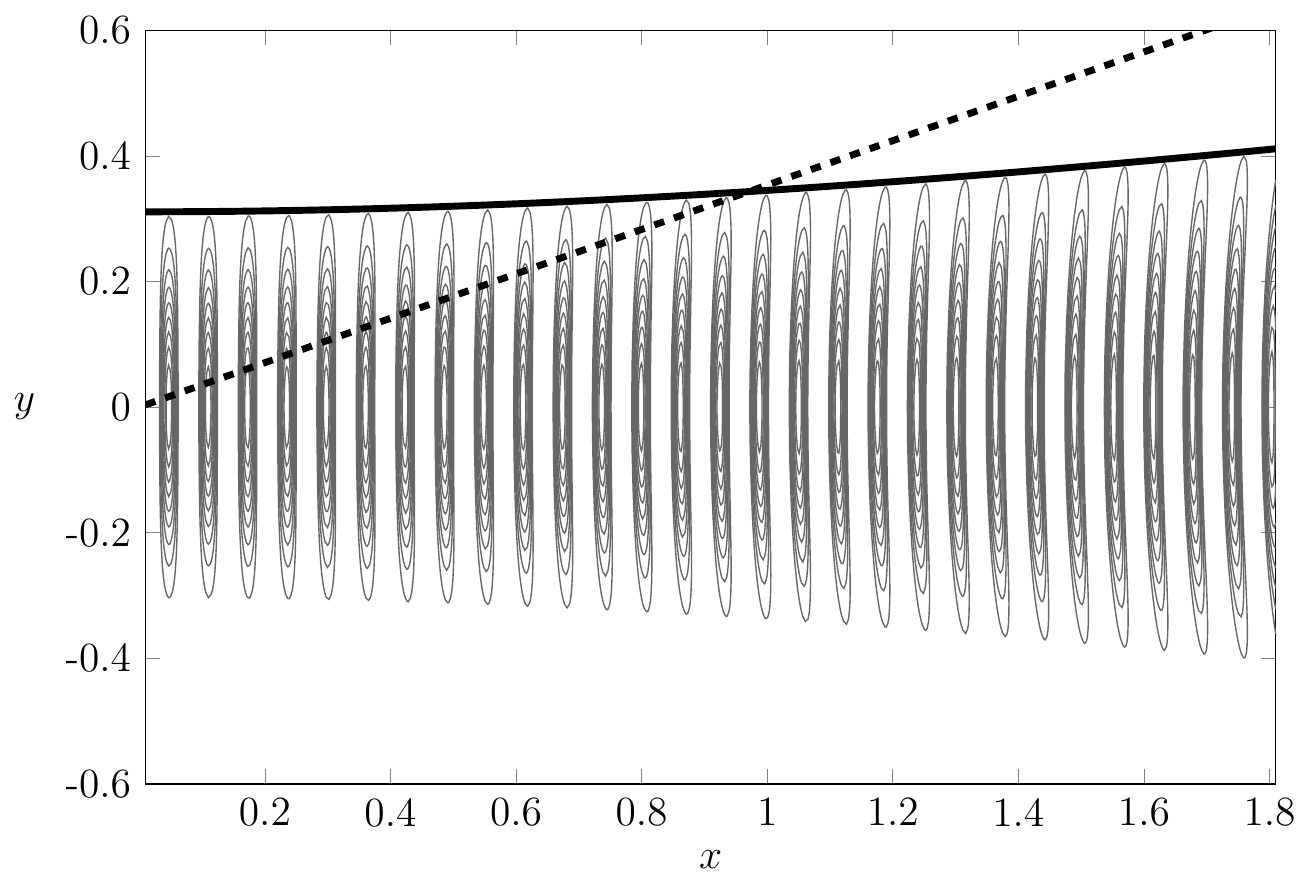}}\hspace{.05\linewidth}
		\subfloat[$F=0.2$]{\includegraphics[width=.45\linewidth]{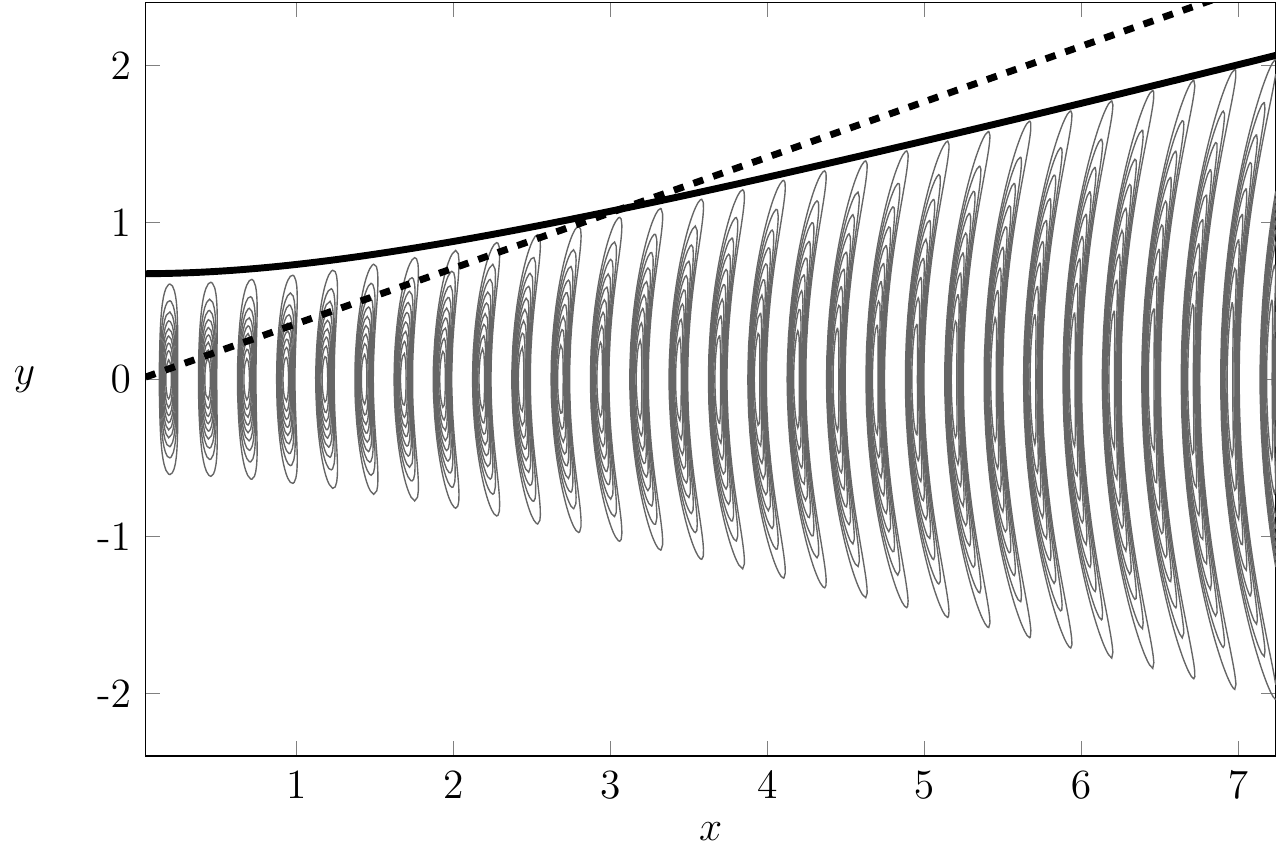}}
		\caption{A contour plot of the positive heights of the free surface profile for flow past a submerged doublet, in which there is a contour for every 10th percentile of the wave height within a single wavelength. The solid line is a line of constant $\mathrm{Re}(\chi)$ as given in (\ref{eq:u}), specifically (a) $\mathrm{Re}(\chi)= 1.0236$ (b) $\mathrm{Re}(\chi)= 1.1026$. The dashed line is Kelvin's angle $\theta_{\mathrm{wedge}}$.}\label{fig:dipoleContour}
	\end{figure}

	We emphasise that the analysis in \cite{lustri13} was performed for the problem of linearised flow past a submerged point source, using the condition (\ref{eq:sourcecond}).  However, the analysis used to derive the governing equation for $\chi$ (\ref{eq:SingulantEq}) and the boundary condition  (\ref{eq:SingulantBC}) does not depend on the behaviour or strength of the singularity at $z=-1$, but only its location.  Therefore, the expression for $\chi$ given in (\ref{eq:u}) may be applied in an identical fashion to derive an envelope function for the analogous problem of flow past a submerged doublet (or even a range of other point singularities).  We illustrate this behaviour in Figure \ref{fig:dipoleContour}, where lines of constant $\mathrm{Re}(\chi)$ are overlaid on the positive heights of the free surface profile for flow past a submerged doublet with $F = 0.1$ and $0.2$. We see that these curves provide an accurate representation of the envelope of the apparent wave region on the surface.

	{
	\subsection{Interference effects at low Froude numbers}
	
	\begin{figure}
		\centering
		\subfloat[$\ell=1.89$]{\includegraphics[width=.32\linewidth]{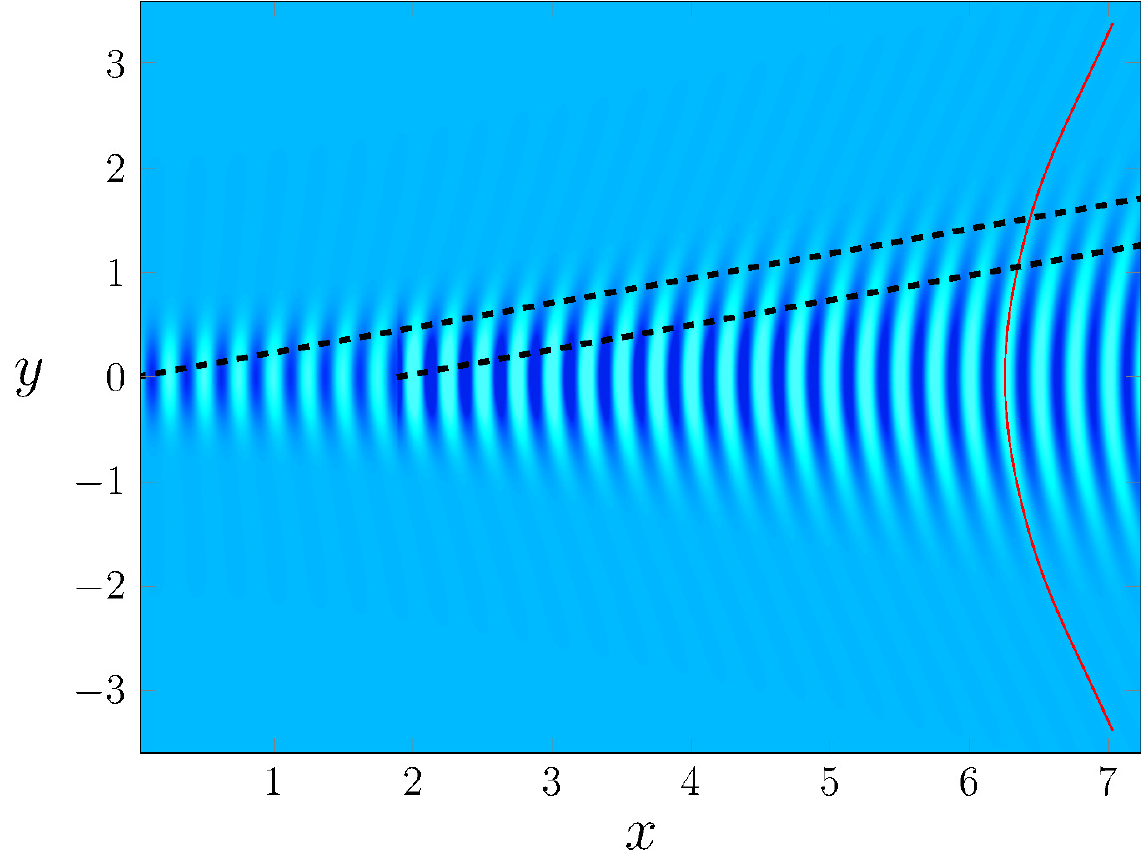}}
		\subfloat[$\ell=1.97$]{\includegraphics[width=.32\linewidth]{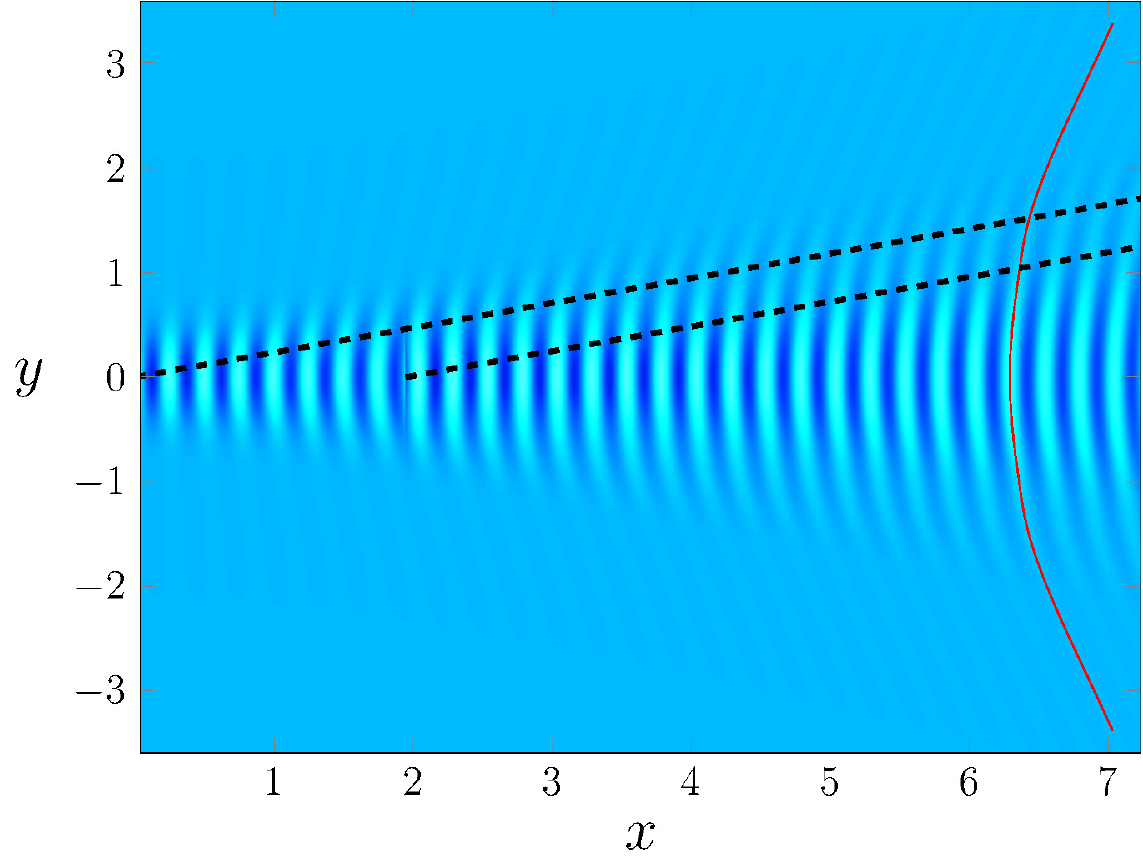}}
		\subfloat[$\ell=2.01$]{\includegraphics[width=.32\linewidth]{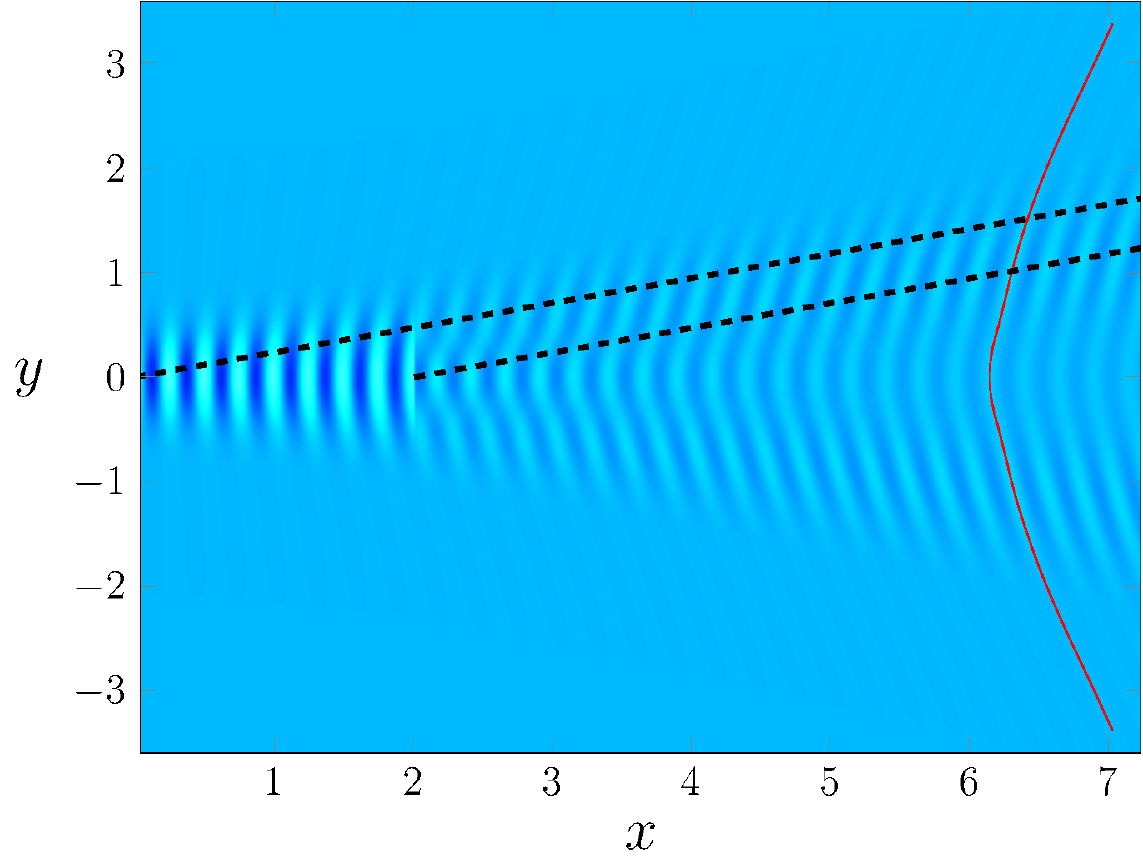}}\\
		\subfloat[Wave crest profiles]{\includegraphics[width=.8\linewidth]{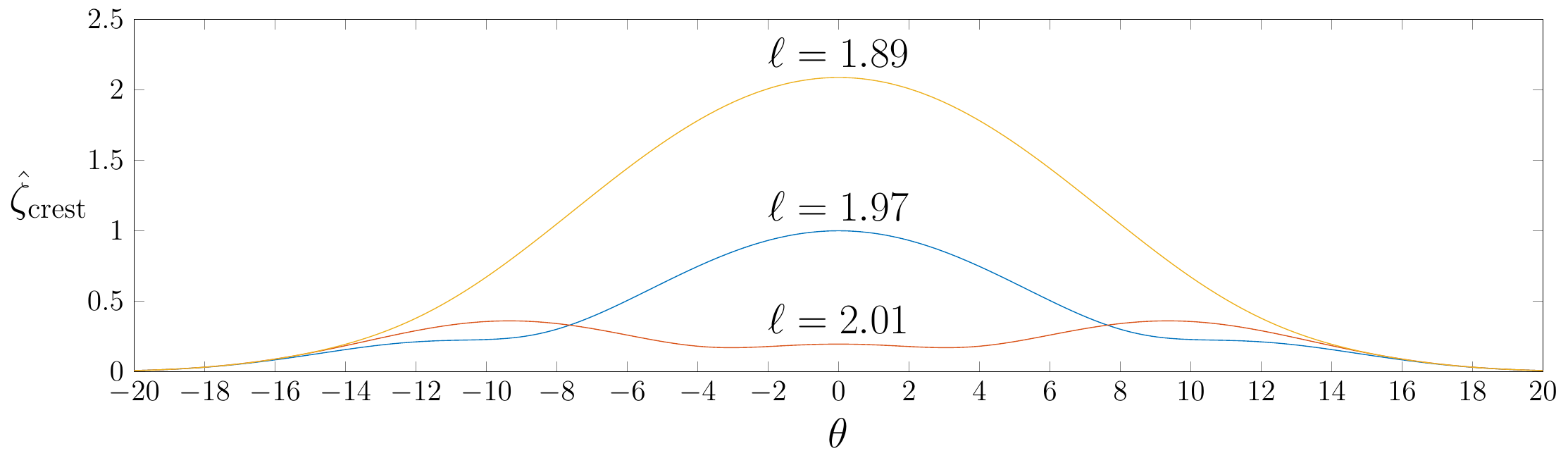}}
		\caption{{ (a)--(c) Plan view of the wave pattern for linear flow past a submerged source and sink, of equal strengths, separated a nondimensional horizontal distance of (a) $\ell=1.89$, (b) $\ell=1.97$, and (c) $\ell=2.01$ for Froude number $F=0.2$.  A superposition of the single integral in (\ref{eq:sourcefunc}) is used for the computation.  The dashed line indicates the wake angle defined by $\alpha=0.2$ for each of the point singularities. The surface height along the red curves in (a)--(c) are plotted in (d) against ray angle $\theta$.}}\label{fig:interferencePlots}
	\end{figure}
	When analysing the apparent wake angle for ship wakes with large Froude numbers, \citet{noblesse14,zhang15a}  have shown that interference between bow and stern waves have the effect of changing the large Froude number scaling in this limit.  Here we will briefly address the effects of interference on the apparent wake angle in the low Froude number regime.  In Figure~\ref{fig:interferencePlots}(a)-(c) we present surface profiles for flow past a Rankine body constructed by a submerging a point source and point sink of equal strengths, where the sink is downstream a horizontal distance $\ell$ from the source (here the dimensionless depth of the source and sink is unity).  In this way, the bow-like waves created by flow past the submerged source may interfere with the stern-like waves due to flow past the submerged sink.  The solutions we present in Figure~\ref{fig:interferencePlots}, generated from the superposition of surface profiles given by (\ref{eq:sourcefunc}) with $F=0.2$, are shown for separation distances (a) $\ell=1.89$, (b) $\ell=1.97$, and (c) $\ell=2.01$,  representing maximum constructive, intermediate, and destructive interference, respectively.

The values of $\ell$ chosen for Figure~\ref{fig:interferencePlots} come from noting that the wavelength of a transverse wave is $2\pi F^2$; for destructive interference we require that $\ell/2\pi F^2$ is an integer, while for constructive interference the same quotient is a half-integer. The intermediate interference profile was chosen such that the quotient, $\ell/2\pi F^2$, has a remainder of $\pm1/6$.  Note the associated length based Froude numbers $F_\ell=F/\sqrt{\ell}$ are (a) $F_\ell=0.1457$, (b) $F_\ell=0.1425$ and (c) $F_\ell=0.1410$.  Using colour intensity to denote wave amplitude, it is easy to see from this figure that even though each of the three geometries is very similar, the effects of wave interference on wave amplitude can be significant.  Indeed, the waves in (a) appear to have a much larger amplitude than in (c).  On the other hand, the apparent wake angle does not appear to change from panels (a)--(c).

To better understand this comparison, we have also presented in Figure~\ref{fig:interferencePlots}(d) the surface elevation along crests marked by the red curves in Figures~\ref{fig:interferencePlots}(a)--(c).  In this figure we have normalised the amplitudes so that the intermediate wave in (b) has a height of unity at $\theta=0$.  It can be readily seen in Figure \ref{fig:interferencePlots}(d) that the maximum elevation of a wave crest varies greatly depending if there is constructive or destructive interference between the bow-lie and stern-like waves. The different maximum elevations mean Method I will give very different apparent wake angle measurements, depending on what type of interference is occurring. Therefore, using Method I to measure the apparent wake angle will no longer give a monotonically decreasing wake angle when Froude number decreases because the surface profile will cycle through constructive and destructive interference as the Froude number changes.
	
	While the maximum height between the three crests in Figure \ref{fig:interferencePlots}(d) vary, all three crest collapse onto the same curve as the ray angle moves away from the centreline. This is because the downstream sink only effects the surface profile within a wedge that is contained wholly with then a wedge that defines the influence of the submerged source. As can be seen in in Figures \ref{fig:interferencePlots}(a)--(c), between the dashed lines is a region on the outer edge of the wake pattern for which only the upstream source contributes to the surface profile. Therefore, even though Method I will give different results between a Rankine body and a single submerged source, the wake angle measurement for a source is applicable for a Rankine body provided its length is sufficiently large.
	
	
	}
	
	\section{Rigid lid approximations}
	\label{sec:slowlymovingobject}
	
	\subsection{Submerged point source ($F\rightarrow 0$ with $\epsilon$ fixed)}\label{sec:rigidlidsource}
	
	Taking the limit of $F\rightarrow 0$ with $\epsilon$ fixed is analogous to flow past a submerged Rankine body with a rounded nose (that approaches a cylinder of radius $\sqrt{\epsilon/\pi}$ in the far field) while reducing the speed of the base flow to zero.  In order to keep the dimensionless parameter $\epsilon$ constant, the dimensional source strength $m$ must also be decreased such that the ratio $m/U$ is constant, thus in this interpretation the source is being turned off as its speed is slowed.
	
	
	In the asymptotic regime $F\rightarrow 0$ with $\epsilon$ fixed, we can approximate the free-surface profile by performing a simple rigid lid approximation similar to that provided in \citet{forbes90}.  To this end, we write
	\begin{equation}
	\phi = \phi_0(x,y,z)+\mathcal{O}(F^2),\quad
	\zeta = F^2\zeta_1(x,y)+\mathcal{O}(F^4)\label{eq:zetaApprox}
	\end{equation}
	and perturb $\phi$ about $z=0$ to give
	\begin{equation}
	\phi(x,y,F^2\zeta_1+\mathcal{O}(F^4))= \phi_0(x,y,0)+F^2\left[\phi_1(x,y,0)+\zeta_1(x,y)\phi_{0z}(x,y,0)\right]+\mathcal{O}(F^4).\label{eq:phiApprox}
	\end{equation}
	We then substitute (\ref{eq:zetaApprox})-(\ref{eq:phiApprox}) into the kinematic and dynamic conditions to derive appropriate boundary conditions for the rigid lid approximation.
	
	The first order term of the kinematic condition gives us $\phi_{0z}(x,y,0)=0$.  Using the method of images with the condition (\ref{eq:sourcecond}), we find
	\begin{equation}
	\phi_0(x,y,z)=x-\frac{\epsilon}{4\pi}\left(\frac{1}{\sqrt{x^2+y^2+(z+1)^2}}+\frac{1}{\sqrt{x^2+y^2+(z-1)^2}}\right).
	\label{eq:phi0}
	\end{equation}
	Taking the first order terms of the dynamic condition and substituting (\ref{eq:phi0}) in yields
	\begin{align}
	\zeta_1(x,y)&=-\frac{\epsilon^2}{8\pi^2}\frac{x^2+y^2}{(x^2+y^2+1)^3}-\frac{\epsilon}{2\pi}\frac{x}{(x^2+y^2+1)^{3/2}}.
	\label{eq:zeta1}
	\end{align}
	From (\ref{eq:zetaApprox}), the leading order approximation is $\zeta=F^2\zeta_1$.
	
\begin{figure}
		\centering
		\subfloat[Nonlinear solution]{\includegraphics[width=.5\linewidth]{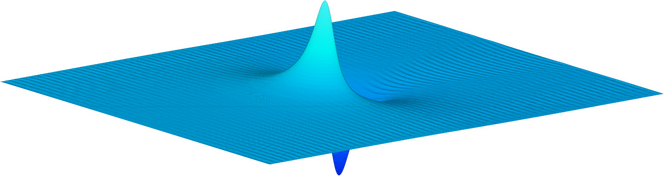}}
		\subfloat[Rigid lid approximation]{\includegraphics[width=.5\linewidth]{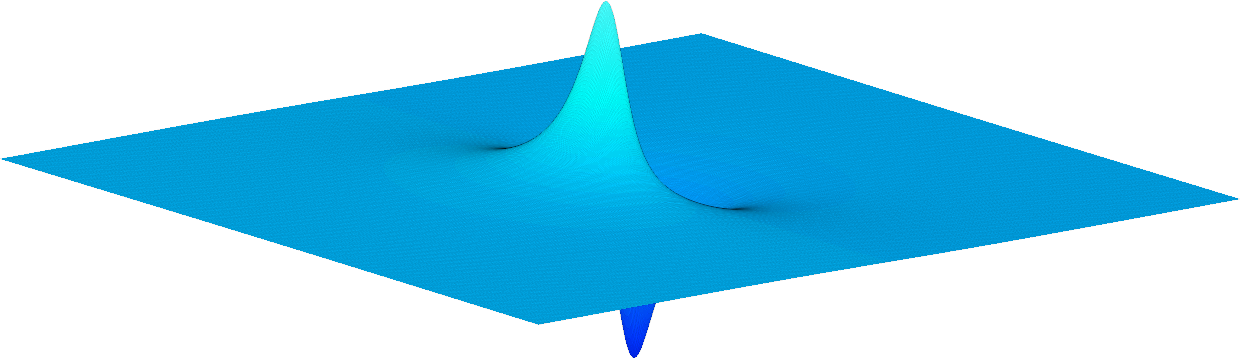}}\\
		\subfloat[Linear solution]{\includegraphics[width=.5\linewidth]{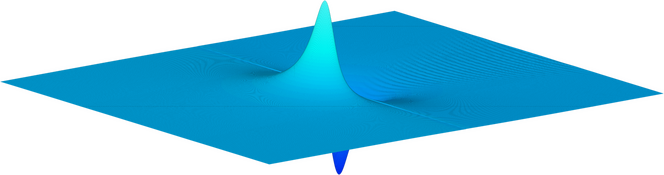}}
		\subfloat[Nonlinear surface minus rigid lid approx.]{\includegraphics[width=.5\linewidth]{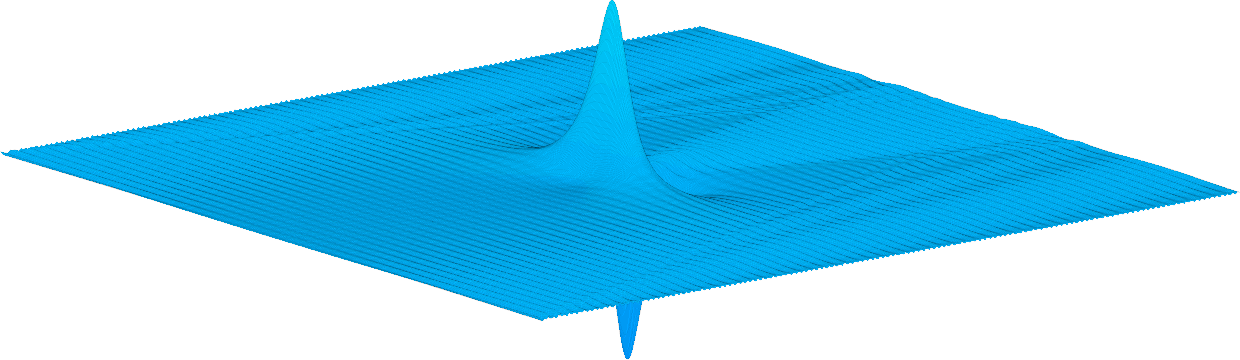}\label{fig:SourceFr03EpSmallMinusRigid}}
		\caption{A comparison of three different free-surface profiles for flow past a point source with $F=0.3$ and $\epsilon=0.1$. The different profiles are of; (a) the full nonlinear solution, (b) the linear solution, (c) the rigid lid approximation and (d) the difference between the nonlinear solution and the rigid lid approximation. In order to give a sense of the scale, the maximum surface heights in these images are (a) $6.27\times 10^{-4}$, (b) $5.50\times 10^{-4}$, (c) $6.27\times 10^{-4}$ and (d) $1.00\times10^{-4}$. The nonlinear solution is presented on a $721\times 241$ mesh with $\Delta x=\Delta y=0.05$ and $x_0=-18$.}\label{fig:SourceSurfRigid}
	\end{figure}

	The rigid lid solution (\ref{eq:zeta1}) is not capable of capturing any of the details of the wave pattern behind the submerged body; however, it does approximate the near field behaviour extremely well, at least for small Froude numbers.  To make such a comparison, we have in Figure~\ref{fig:SourceSurfRigid} representative free-surface profiles for the example $F=0.3$, $\epsilon=0.1$.  In Figure~\ref{fig:SourceSurfRigid}(a), the numerical solution to the fully nonlinear problem (\ref{eq:laplace})--(\ref{eq:farfield}) (with
	$\delta=0$) and (\ref{eq:sourcecond}) is shown, while in (b) we have the rigid lid solution (\ref{eq:zetaApprox}) with (\ref{eq:zeta1}).  On this scale, both surfaces appear to compare very well with each other.
	
	Also included in Figure~\ref{fig:SourceSurfRigid} is the linear solution (\ref{eq:sourcefunc}) from \S\ref{sec:linearsource}.  For these chosen parameters, the linear solution is also very similar to the full numerical solution, although the match is not as good as the rigid lid solution.  The one key difference is that the linear solution is able to capture the small transverse waves, while (as just mentioned) the rigid lid solution is not.
	
	Given the amplitude of the waves downstream for the disturbance is extremely small for this example, it is difficult to extract these waves from the full numerical solution in part (a) of Figure~\ref{fig:SourceSurfRigid}.  We can, however, subtract the rigid lid solution (\ref{eq:zetaApprox}), (\ref{eq:zeta1}) from the numerical solution to isolate these waves, as in Figure \ref{fig:SourceSurfRigid}(d).  Thus the rigid lid solution turns out to be useful to help analyse the full numerical solution for small Froude numbers.
	
	In order to perform a full asymptotic analysis in the limit $F\rightarrow 0$ with $\epsilon$ fixed, we require techniques in exponential asymptotics that have not yet been fully developed.  We discuss this issue in \S\ref{sec:discussion}.
	
	\subsection{Submerged point doublet ($F\rightarrow 0$ with $\mu$ fixed)}\label{sec:rigidliddoublet}
	
	By fixing $\mu$ and taking the limit $F\rightarrow 0$, the configuration tends to a flow of decreasing magnitude past a submerged spherical body (of radius $(\mu/2\pi)^{1/3}$).  In dimensional terms, fixing $\mu$ while decreasing $F$ is equivalent reducing both $U$ and $\kappa$ while keeping $\kappa/U$ constant.  Thus, another interpretation of this limit is the doublet is gradually being turned off as it is being slowed down.
	
	The rigid lid approximation can also be applied to this flow configuration.  Omitting the details, we find
	\begin{align}
	\zeta(x,y)=&\,F^2\left[\frac{\mu^2}{8\pi^2}\left(-\frac{1}{(x^2+y^2+1)^3}+\frac{6x^2}{(x^2+y^2+1)^4}-\frac{9(x^4+x^2y^2)}{(x^2+y^2+1)^5}\right)\right.\notag\\
	&\qquad\left.+\frac{\mu}{2\pi}\left(-\frac{1}{(x^2+y^2+1)^{3/2}}+\frac{3x^2}{(x^2+y^2+1)^{5/2}}\right)\right].
	\label{eq:rigidliddoublet}
	\end{align}
	For brevity we shall not include images of the surface (\ref{eq:rigidliddoublet}); however, it turns out that the results are analogous to Figure~\ref{fig:SourceSurfRigid} in that the rigid lid solution (\ref{eq:rigidliddoublet}) provides a very good approximation for the full nonlinear solution in the near field, but is deficient in the sense that it does not describe the wave train downstream.  Further discussion on this asymptotic limit is deferred until \S\ref{sec:discussion}.
	
	\subsection{Applied pressure distribution ($F\rightarrow 0$ with $\delta$ fixed)}\label{sec:rigidlidpressure}
	
	Here we suppose that $F\rightarrow 0$ with $\delta$ fixed, which has the physical interpretation of turning the pressure off as it is being slowed down.  The rigid lid approximation for this regime is given by the straightforward expression
	\begin{equation}
	\zeta(x,y)=-F^2\delta p(x,y).
	\end{equation}
	Again, to save space we shall not go into any further details here, except to repeat that the rigid lid approximation does a very good job of predicting the near field behaviour for small $F$, but does not capture the wave pattern, as the wave amplitudes are exponentially small compared to powers of $F^2$.
	
	\section{Slowly moving submerged point singularity and pressure distribution}
	\label{sec:slowlymovingsourcedoublet}
	
	\subsection{Submerged point source ($F\rightarrow 0$ with $\epsilon F$ fixed)}\label{sec:slowpoint}
	
\begin{figure}
		\centering
		\subfloat[$F\rightarrow 0$]{\includegraphics[width=0.33\linewidth]{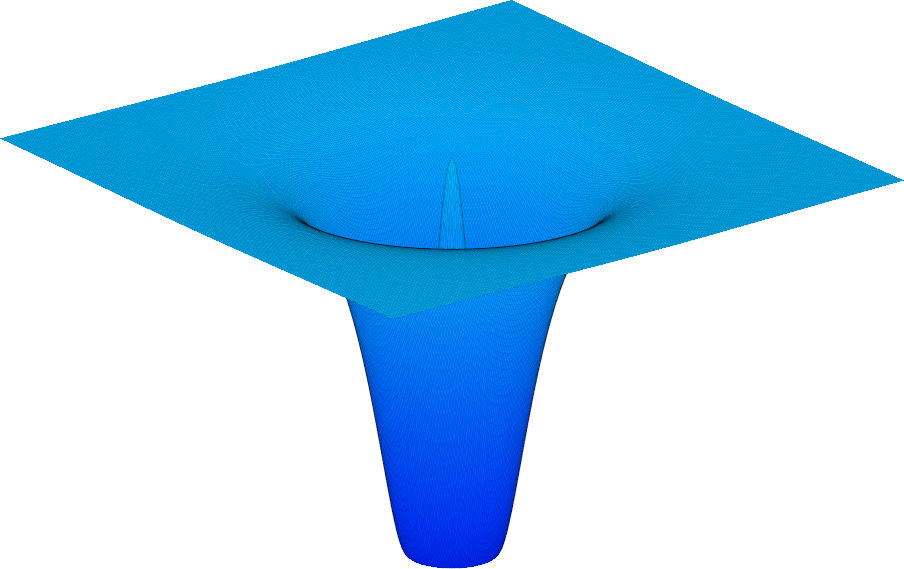}}
		\subfloat[$F=0.03$]{\includegraphics[width=0.33\linewidth]{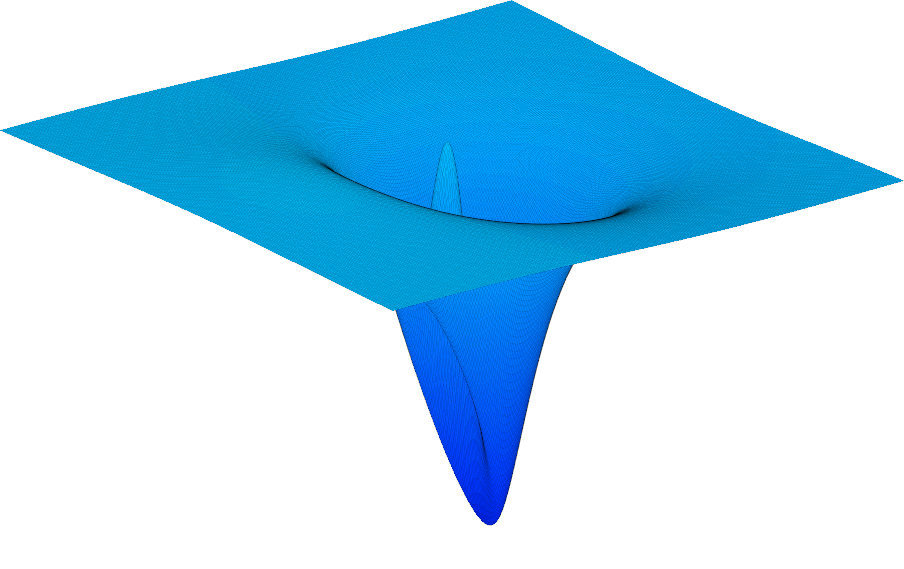}}
		\subfloat[$F=0.07$]{\includegraphics[width=0.33\linewidth]{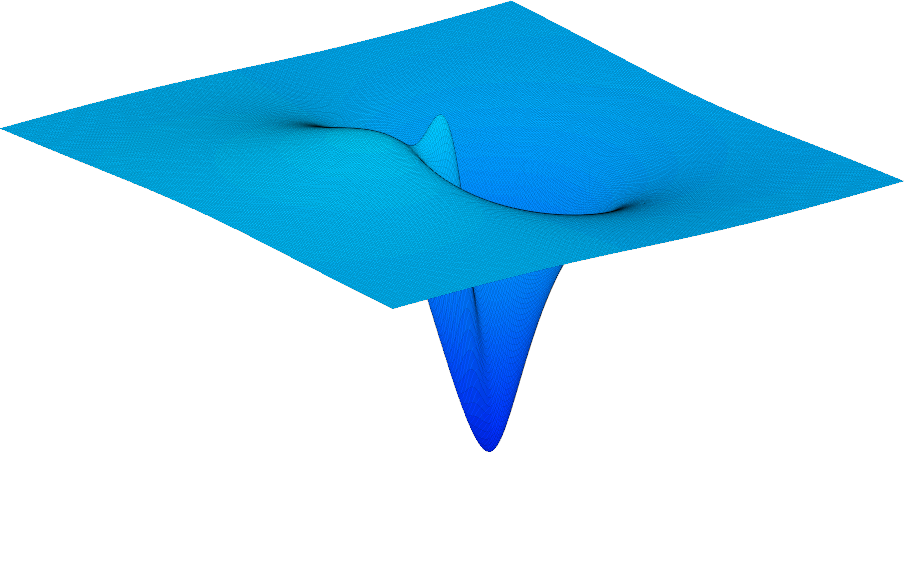}}\\
		\subfloat[$F\rightarrow 0$]{\includegraphics[width=0.33\linewidth]{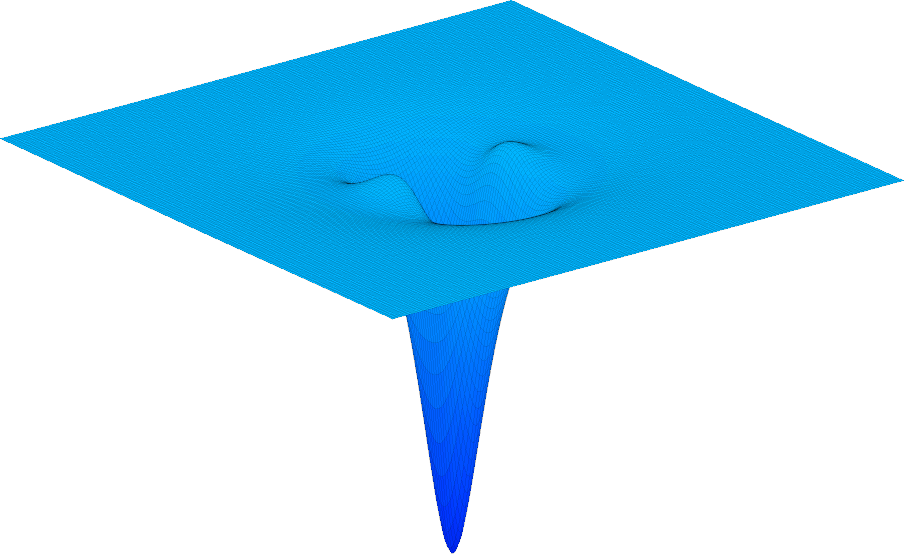}}
		\subfloat[$F=0.02$]{\includegraphics[width=0.33\linewidth]{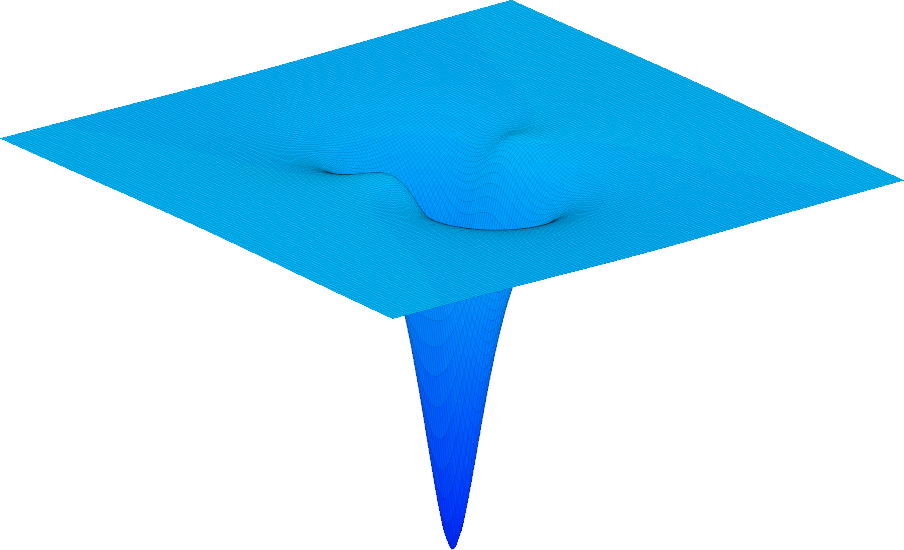}}
		\subfloat[$F=0.04$]{\includegraphics[width=0.33\linewidth]{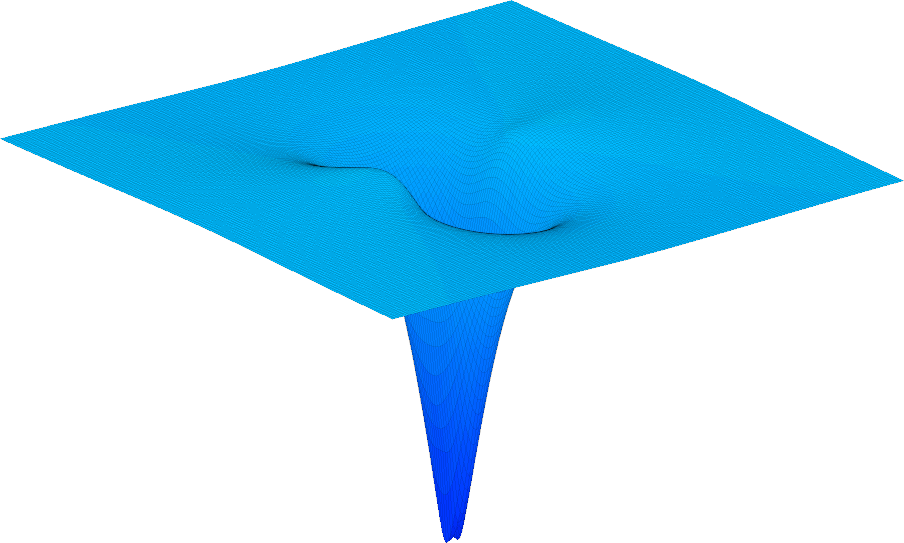}}\\
		\subfloat[$F\rightarrow 0$]{\includegraphics[width=0.33\linewidth]{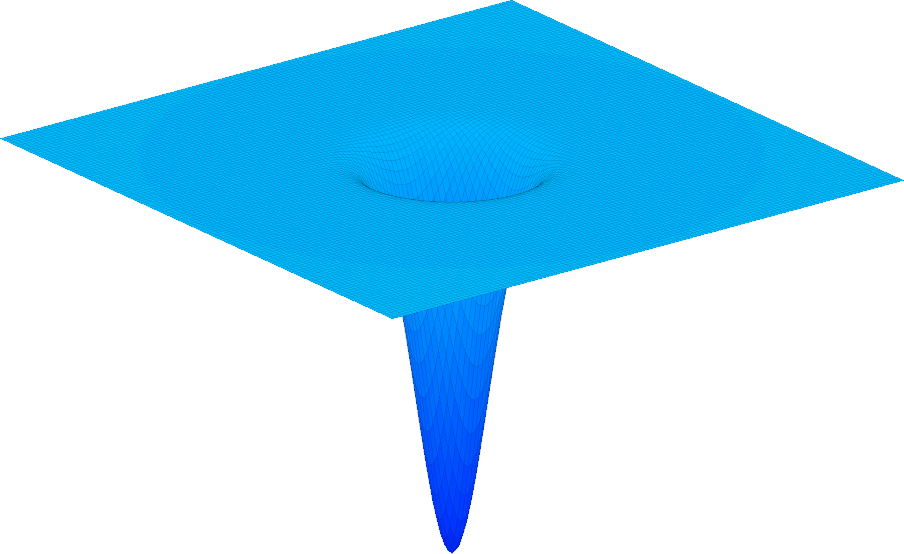}}
		\subfloat[$F=0.06$]{\includegraphics[width=0.33\linewidth]{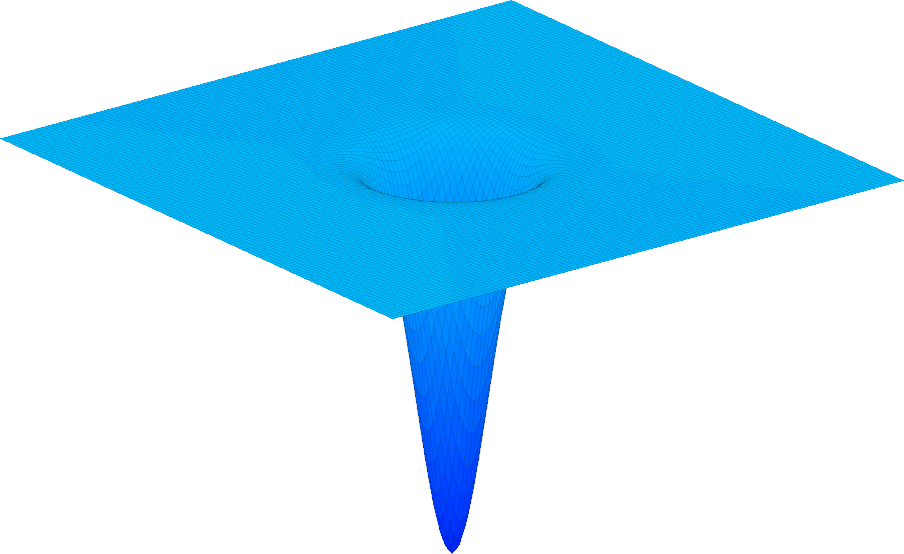}}
		\subfloat[$F=0.12$]{\includegraphics[width=0.33\linewidth]{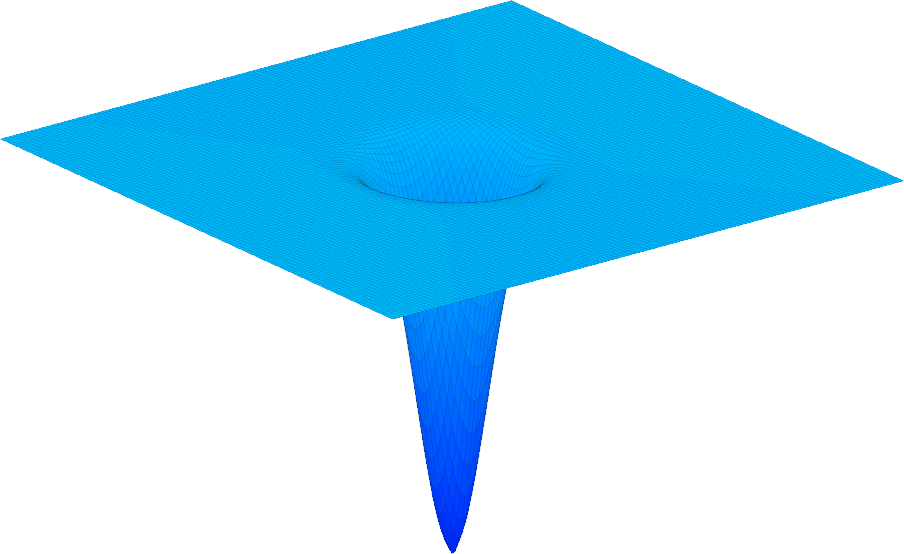}}
		\caption{A free-surface profile for nonlinear flow past: (a)--(c) a point source with $\epsilon F=2$, (d)--(f) a point doublet with $\mu F=1$, and (g)--(i) a pressure distribution with $\delta F^2=3$. All solutions are presented on a $721\times 361$ mesh with $\Delta x=\Delta y=0.033$ and $x_0=-12$.}\label{fig:SlowSource}
	\end{figure}

	In the double limit $F\rightarrow 0$, $\epsilon\rightarrow\infty$ with $\epsilon F$ held constant, the solution approaches that for flow due to a stationary submerged point source explored in \citet{forbes90,vandenbroeck97a,hocking16}, governed by the single parameter $\bar{m}=\epsilon F=m/\sqrt{gL^5}$.  This base flow is itself a highly nontrivial free boundary problem, which these authors demonstrate has solutions up to some limiting value $\bar{m}$, $\bar{m}_{\mathrm{max}}$ say.  There is some uncertainty about what this limiting value is, given the challenges introduced by the nonlinearities involved.  In Figure~\ref{fig:SlowSource}(a), we present an axisymmetric free-surface profile  for this base case for the single value $\bar{m}=2$.  This is a highly nonlinear solution which is close the limiting configuration for $\bar{m}=\bar{m}_{\mathrm{max}}$.
	
	To illustrate the regime with $F\ll 1$, $\bar{m}$ constant, we include two nonlinear free-surface profiles in Figure~\ref{fig:SlowSource}(b)-(c) with $\bar{m}=2$, this time for $F=0.03$ ($\epsilon=200/3$) and $F=0.07$ ($\epsilon=200/7$).  These are clearly non-axisymmetric surfaces which appear to be small perturbations of the solution in Figure~\ref{fig:SlowSource}(a).  They correspond to slowly moving submerged point sources.

{Note that in Figure~\ref{fig:SlowSource}(a) there is a central peak in the surface elevation at the origin.  This peak is accompanied by a stagnation point which must occur on the surface directly above the source due to symmetry. In panels (b) and (c), for which there is a nonzero flow in the far field, we see that the central peak has moved slightly along the $x$-axis in the negative direction.  In these two cases, the height of the peaks is less than $F^2/2$ and so these peaks are not stagnation points.}

	\subsection{Submerged point doublet ($F\rightarrow 0$ with $\mu F$ fixed)}\label{sec:slowdoublet}
	
	If we take $F\rightarrow 0$ and $\mu\rightarrow\infty$ with $\mu F=\bar{\mu}$ held constant, then the limiting flow is that due to a stationary point doublet.  As far as we can tell, the fully nonlinear version of this configuration does not appear to be considered in the literature.  We have not undertaken a detailed analysis of the parameter space for this interesting problem, as it is outside the scope of the present study. Instead, we present a representative profile in Figure~\ref{fig:SlowSource}(d) for $\bar{\mu}=1$.  Then in (e) and (f) we show solutions with the same value of $\bar{\mu}=1$, but with $F=0.02$ and $0.04$, respectively.  These profiles with $F>0$ are clearly small perturbations of the base configuration with a stationary point doublet.  They correspond to a slowly moving submerged point doublet.
	
	\subsection{Applied pressure distribution ($F\rightarrow 0$ with $\delta F$ fixed)}\label{sec:slowpressure}
	
	In the limit $F\rightarrow 0$, $\delta\rightarrow\infty$ with $\delta F^2=\bar{\delta}$ fixed, the flow approaches that due to a stationary pressure distribution.  An example of this trivial configuration is shown in Figure~\ref{fig:SlowSource}(g), where $\bar{\delta}=3$.  We can perturb this flow by moving the applied pressure distribution very slowly, for example as in Figure~\ref{fig:SlowSource}(h)-(i).  These two rather unremarkable solutions are also for $\bar{\delta}=3$, except this time $F=0.06$ and $0.12$, respectively.
	
	\section{Highly nonlinear regime}
	\label{sec:nonlinearflows}
	
	In this section we fix the Froude number $F$ to be small and apply our numerical scheme to increase $\epsilon$, $\mu$ or $\delta$ as much as possible. This is a highly nonlinear regime which, as far as we can tell, has not received any attention (at least for three-dimensional flows).  We are able to compute distinctive wave patterns that have highly nonlinear features not observed in any other parameter regime.
	
	\subsection{Submerged point source ($\epsilon\rightarrow \epsilon_{\mathrm{max}}$ with $F$ fixed)}\label{sec:nonlinearpoint}
	
	To begin, consider the problem of flow past a submerged point source.  By setting $F=0.3$, $\epsilon=0.1$, we can compute a solution which is only weakly nonlinear, as shown in Figure~\ref{fig:SourceSurf}(a).  Now as $\epsilon$ increases, the effects of nonlinearity are more prominent.  A train of waves appears behind the disturbance.  If we keep increasing $\epsilon$, then we can attempt to explore the limit $\epsilon\rightarrow \epsilon_{\mathrm{max}}$ with $F$ fixed, where $\epsilon_{\mathrm{max}}$ is defined such that the free surface $\zeta(x,y)$ for $\epsilon=\epsilon_{\mathrm{max}}$ has the property $\max(\zeta)=F^2/2$ (as permitted by Bernoulli's equation (\ref{eq:dyncond})). This would be the limiting configuration characterised by a stagnation point where $\zeta(x,y)=\max(\zeta)$.  With our numerical scheme, we are not able to compute the actual limiting configuration, but we are able to resolve solutions that are close to the limit.  For example, with $F=0.3$, the largest value of $\epsilon$ we were able to compute a solution was $\epsilon=8.5$.  The corresponding free-surface profile is shown in Figure~\ref{fig:SourceSurf}(b). { As an indication of how nonlinear this solution is, the maximum slope of the surface along the centreline is $32^\circ$, which occurs immediately before the highest peak.  It is interesting to note this steepness is slightly higher than that for the steepest possible two-dimensional wave ($30^\circ$) \citep{schwartz74}, known as the Stokes limiting configuration \citep{stokes1847}.}
	
	Looking closely at Figures~\ref{fig:SourceSurf}(a)-(b), we see there is an initial peak in the free-surface profile.  This initial peak is the only prominent feature in Figure~\ref{fig:SourceSurf}(a), for which the strength of the source is small.  On the other hand, in Figure~\ref{fig:SourceSurf}(b), at the highest source strength we could solve for, there is also the initial peak, but this is followed by many further peaks along the wave train. It turns out that the first peak of the wave train is higher than the initial peak in Figure~\ref{fig:SourceSurf}(b).    If we assume the solution will reach its limiting configuration when $\zeta(x,y)=\max(\zeta)$ at one of the peaks, it is of interest to speculate which peak is the cause of the solution breaking down as $\epsilon\rightarrow \epsilon_{\mathrm{max}}$.
	
	We do this in Figure~\ref{fig:CentrelinePeaks}(a)-(b) for two values of the Froude number, where we plot the height of the initial peak and the first peak of the wave train against the strength of the source, $\epsilon$, remembering that the higher the value of $\epsilon$ the more nonlinear the solution is.  We also plot the limiting height $F^2/2$ as a horizontal dashed line, remembering that no solutions exist with any value of $\zeta>\max(\zeta)=F^2/2$.  For $F=0.2$, in Figure~\ref{fig:CentrelinePeaks}(a) we see the blue data points representing the initial peak are much higher than the first peak of the wave train, and so it is almost certain that the initial peak is approaching the limiting height as $\epsilon\rightarrow \epsilon_{\mathrm{max}}$.  In contrast, for $F=0.3$ the pattern appears the same for small and moderate values of $\epsilon$, with the initial peak much higher than the first peak of the wave train, but then for a very large value of $\epsilon$ the data intersects.  Therefore, for $F=0.3$ it appears that the first peak of the wave train is the cause of the solution reaching the limiting configuration.  As far as we can tell, while this type of issue is explored in some detail for steady flows in two dimensions, the more complicated analogues in three dimensions are relatively unexplored in the literature \citep{buttle18,pethiyagoda14b}.

	\subsection{Submerged point doublet ($\mu\rightarrow \mu_{\mathrm{max}}$ with $F$ fixed)}\label{sec:nonlineardoublet}
	
	For the problem of flow past a submerged point doublet, there are some notable differences when compared to the previous subsection.  First, we show a weakly nonlinear solution in Figure~\ref{fig:SourceSurf}(c) for $\mu=0.1$ and a highly nonlinear solution in Figure~\ref{fig:SourceSurf}(d) for $\mu=1.58$.  While there are some relatively small waves behind the disturbance for $\mu=0.1$, these waves dominate the wake for $\mu=1.58$.  The solution presented in Figure~\ref{fig:SourceSurf}(d) is the most nonlinear solution we could compute in the sense that $\mu=1.58$ is the largest value of the doublet strength for which our numerical scheme would converge.  { These waves here are very steep, with the maximum slope along the centreline roughly $25^\circ$.}  A clear qualitative difference between this highly nonlinear solution and the one in Figure~\ref{fig:SourceSurf}(b) for flow past a point source is that the wave pattern in Figure~\ref{fig:SourceSurf}(d) is characterised by very steep divergent waves.  This is surprising as the Froude number here is very low and a rule of thumb is that ship wave profiles for small Froude number are dominated by transverse waves.
	
	Another observation from Figure~\ref{fig:SourceSurf}(c)-(d) is that in both cases we see the initial peak of the free-surface profile is lower than the first peak of the wave train, and in the highly nonlinear case in Figure~\ref{fig:SourceSurf}(d) the difference in heights is substantial.  This means that, ultimately, the solution will reach a limiting configuration as $\mu\rightarrow \mu_{\mathrm{max}}$ when the height of the first peak of the wave train reaches $\max(\zeta)=F^2/2$.  To demonstrate how these two peaks depend on the strength of the doublet, we show provide data in Figure~\ref{fig:CentrelinePeaks}(c)-(d) for two values of the Froude number $F=0.2$ and $0.3$.  The qualitative difference between Figure~\ref{fig:CentrelinePeaks}(c)-(d) and Figure~\ref{fig:CentrelinePeaks}(a)-(b) is clear.
	
	\subsection{Applied pressure distribution ($\delta\rightarrow \delta_{\mathrm{max}}$ with $F$ fixed)}\label{sec:nonlinearpressure}
	
\begin{figure}
		\centering
		\subfloat[$\epsilon=0.1$]{\includegraphics[width=.5\linewidth]{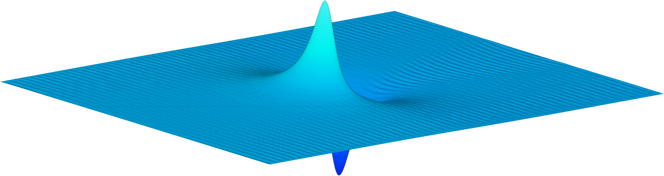}}
		\subfloat[$\epsilon=8.5$]{\includegraphics[width=.5\linewidth]{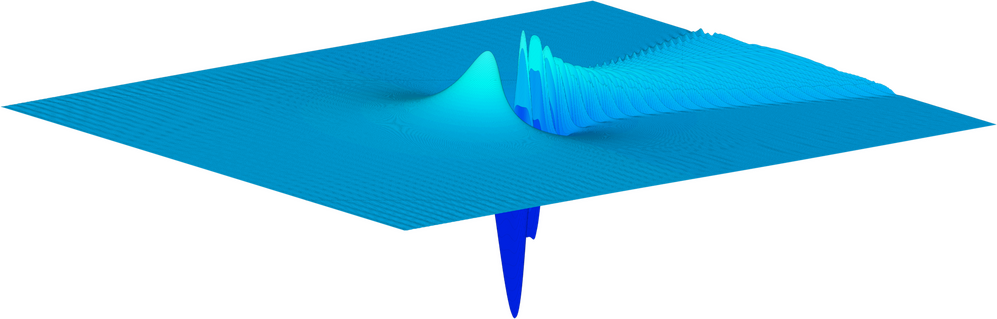}}\\
		\subfloat[$\mu=0.1$]{\includegraphics[width=.5\linewidth]{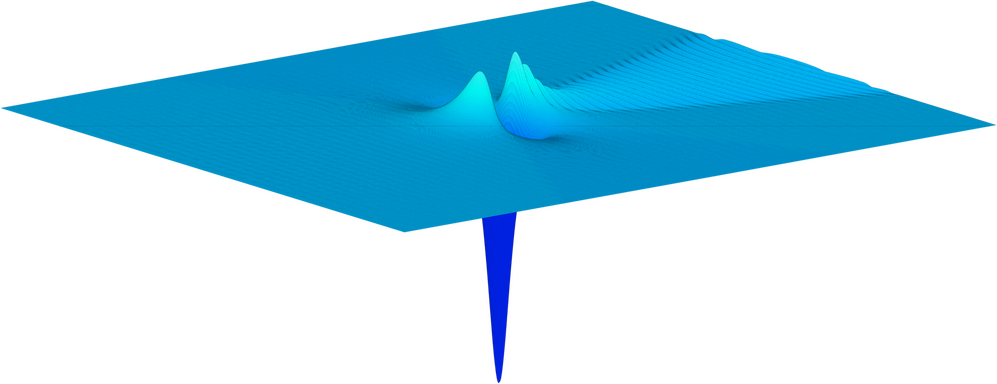}}
		\subfloat[$\mu=1.58$]{\includegraphics[width=.5\linewidth]{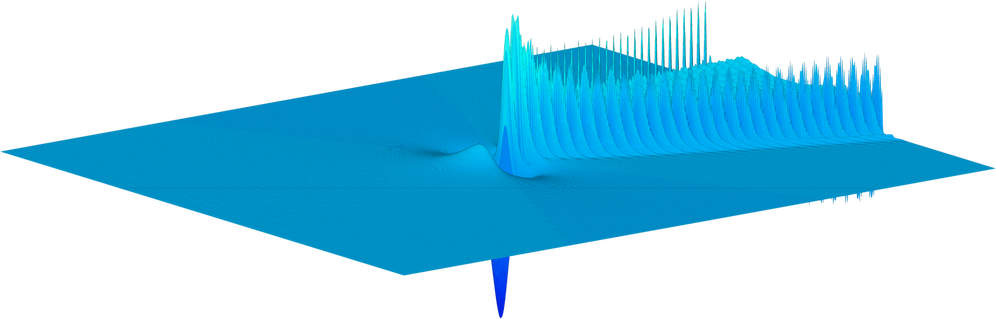}}\\
		\subfloat[$\delta=0.1$]{\includegraphics[width=.5\linewidth]{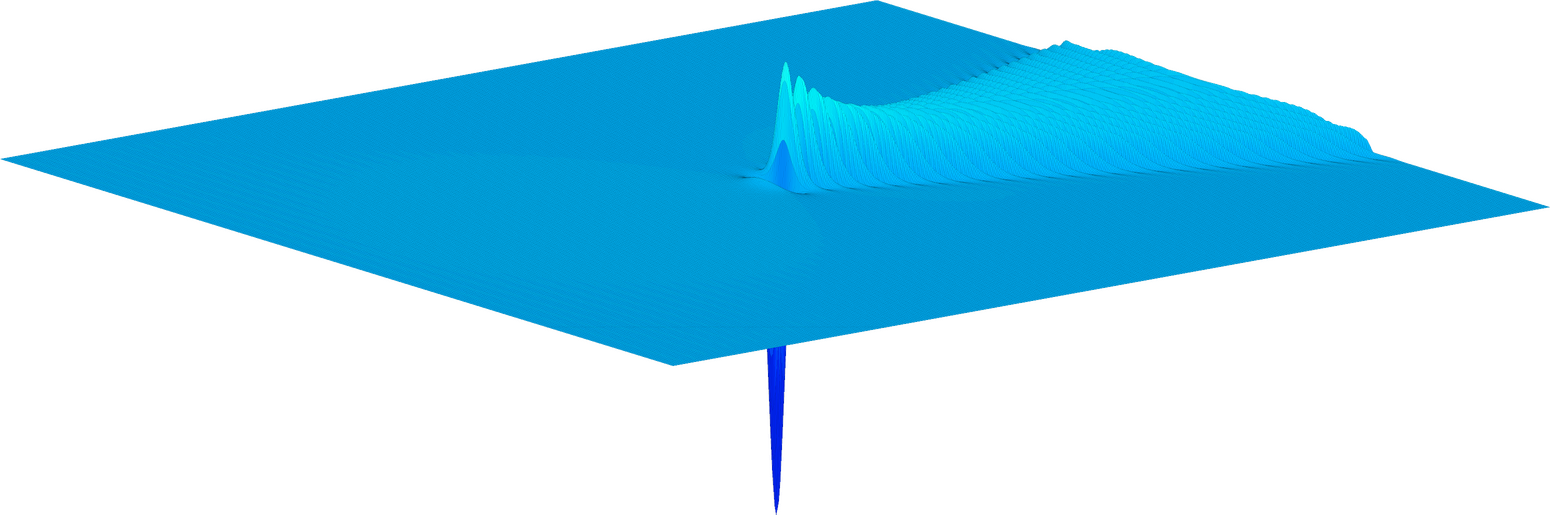}}
		\subfloat[$\delta=0.55$]{\includegraphics[width=.5\linewidth]{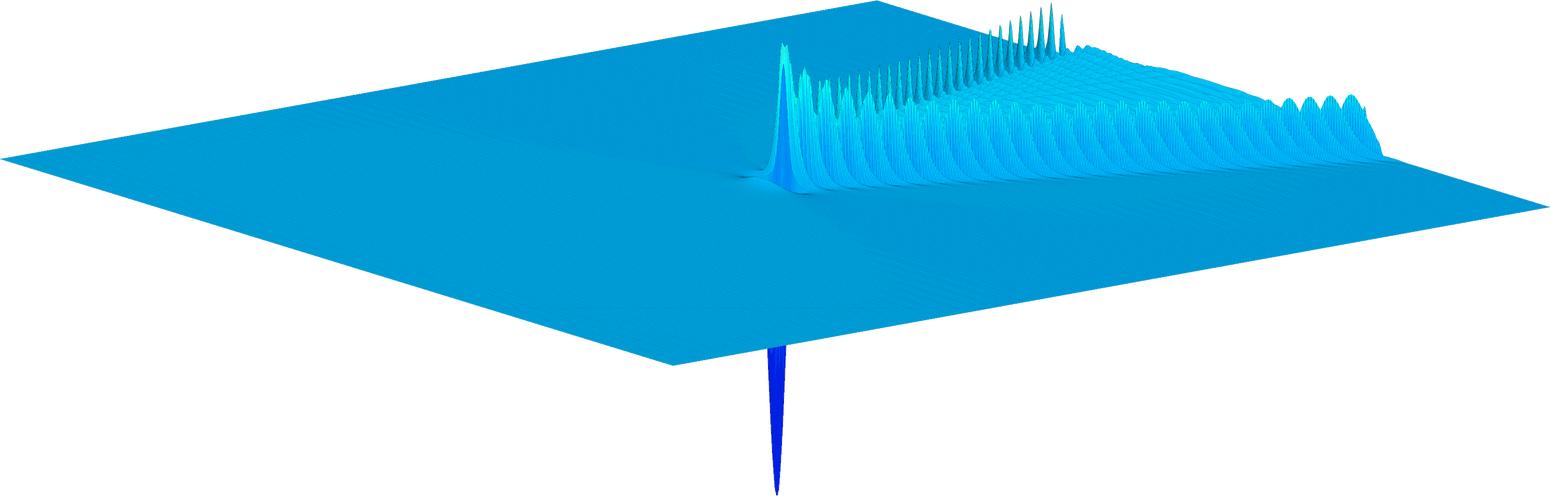}}
		\caption{Free-surface profiles for flow past (a,b) a submerged point source, (c,d) a submerged point doublet, and (e,f) an applied pressure distribution for Froude number $F=0.3$. Both weakly nonlinear and (a,c,e) highly nonlinear (b,d,f) solutions shown. {The maximum gradients of the surface elevation along the centreline for each case is roughly (a) $0.02^\circ$, (b) $32^\circ$, (c) $0.3^\circ$, (d) $25^\circ$, (e) $5^\circ$, and (f) $35^\circ$.} All solutions are presented on a $721\times 241$ mesh with $\Delta x=\Delta y=0.05$ and $x_0=-18$.}\label{fig:SourceSurf}
	\end{figure}

	We have also studied the problem of flow past a pressure distribution by fixing the Froude number $F$ and increasing the strength of the pressure $\delta$.  We compare two free-surface profiles in Figure~\ref{fig:SourceSurf}(e)-(f).  The solution in (e) is only weakly nonlinear as $\delta$ is rather small, while the solution in (f) is highly nonlinear{ with maximum slope along the centreline of roughly $35^\circ$.}  As with Figure~\ref{fig:SourceSurf}(d), we see in Figure~\ref{fig:SourceSurf}(f) that the wave pattern is dominated by the divergent waves, which is surprising since the Froude number is small and we may expect the divergent waves to be much smaller than the transverse waves in this regime.  The strong nonlinearity here is obviously causing this counterintuitive behaviour.
	
\begin{figure}
		\centering
		\subfloat[Source, $F=0.2$]{\includegraphics[height=.33\linewidth]{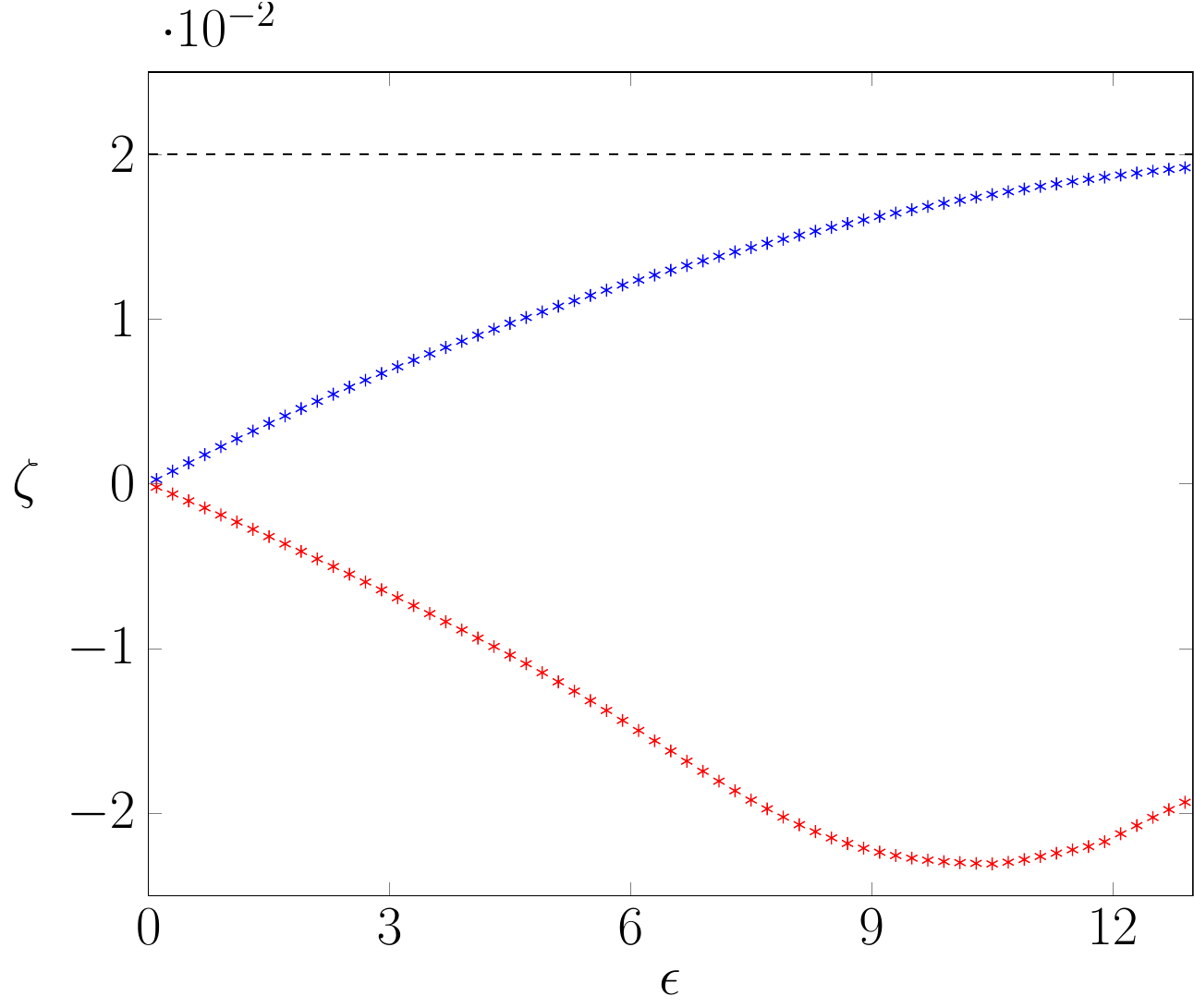}}\hspace{2ex}
		\subfloat[Source, $F=0.3$]{\includegraphics[height=.33\linewidth]{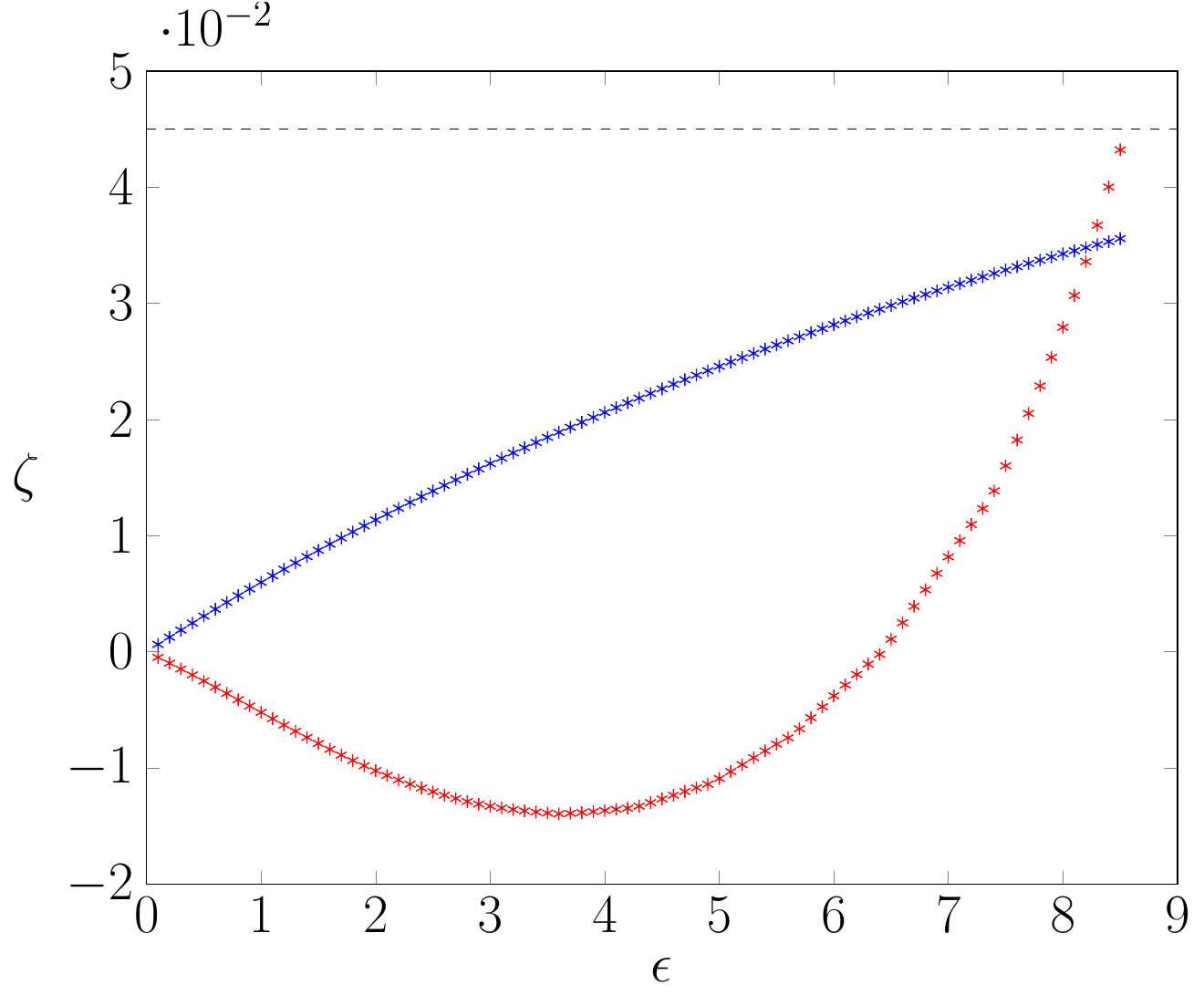}}\\
		\subfloat[Doublet, $F=0.2$]{\includegraphics[height=.33\linewidth]{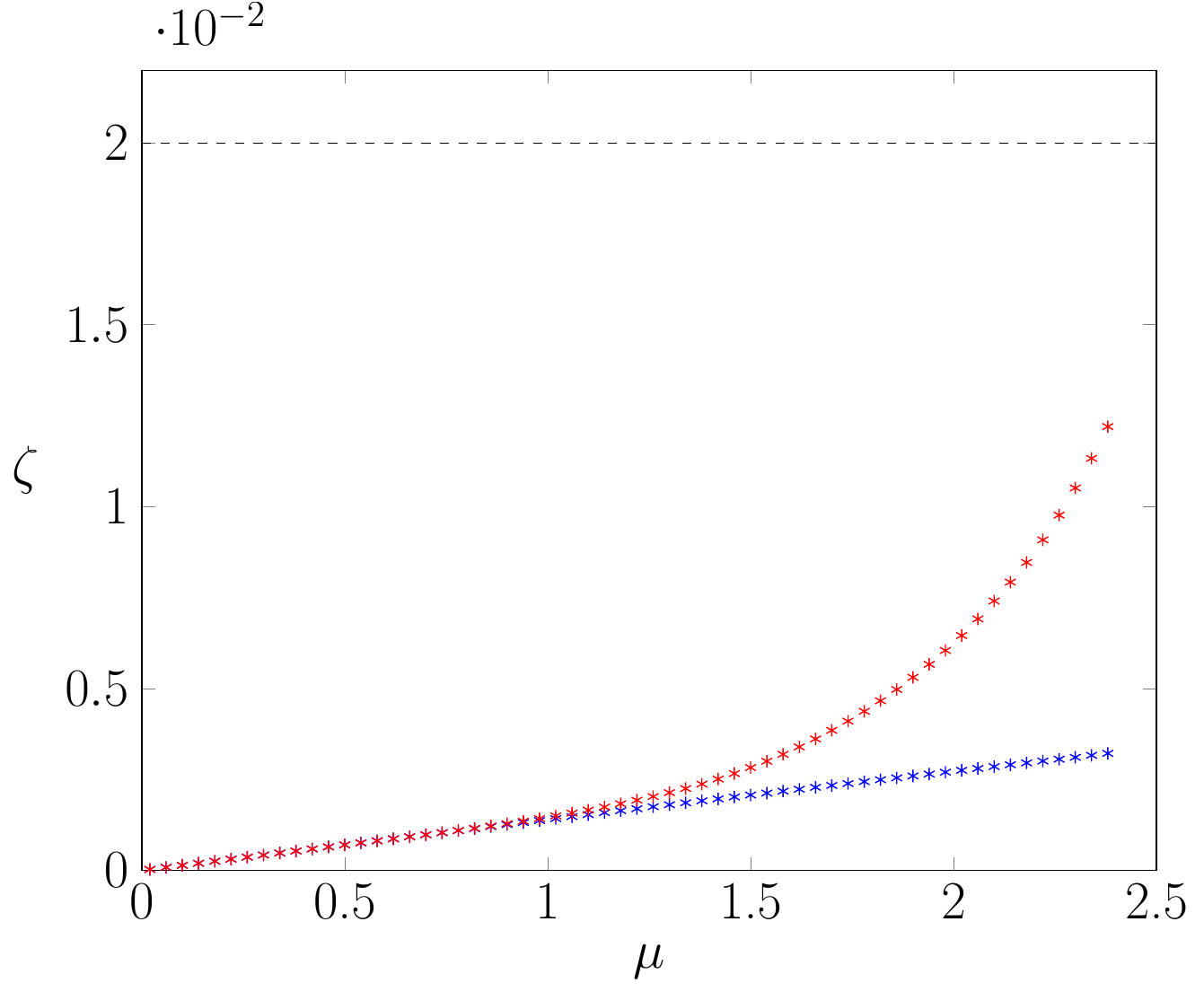}}\hspace{1.5ex}
		\subfloat[Doublet, $F=0.3$]{\includegraphics[height=.33\linewidth]{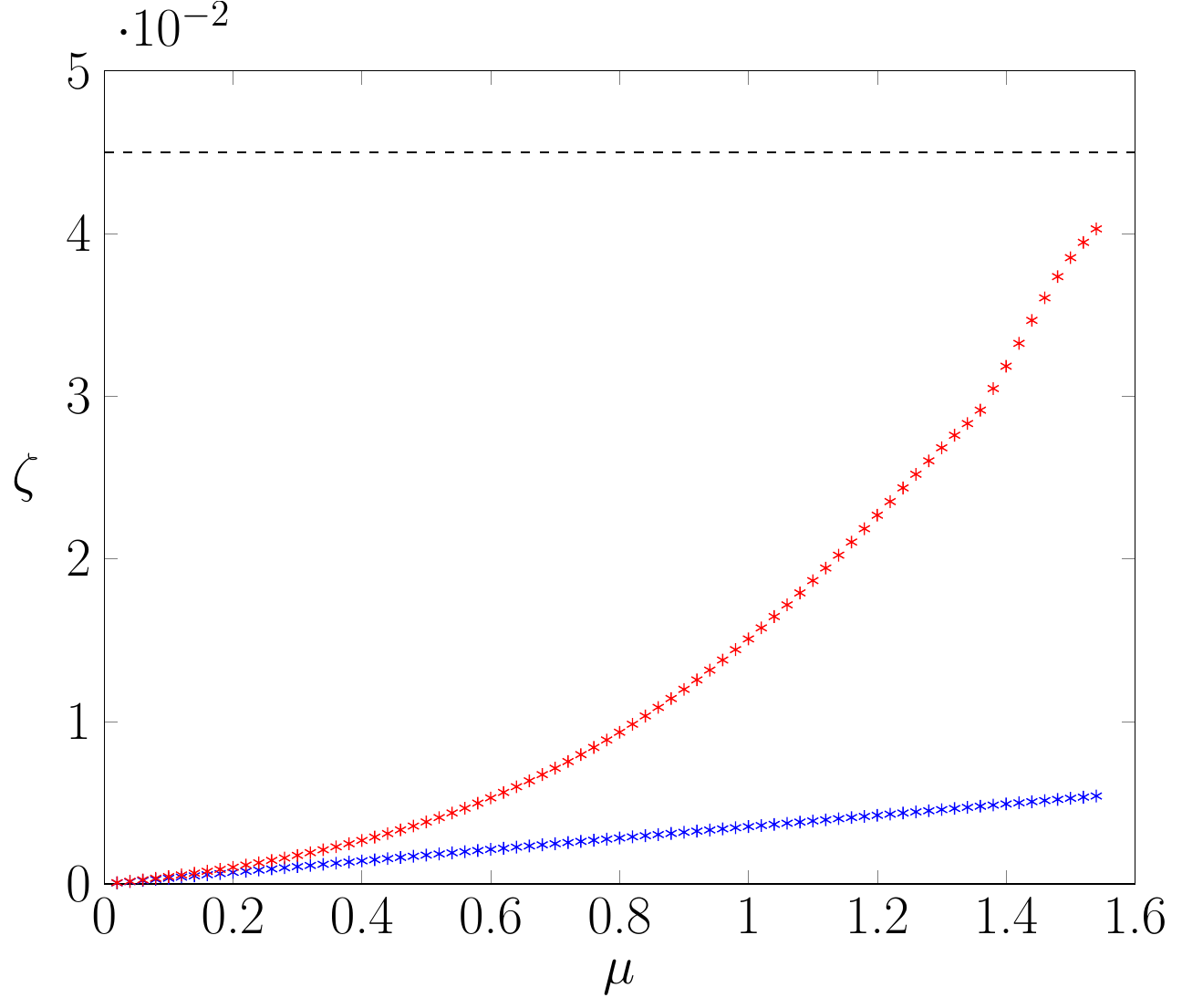}}\\
		\subfloat[Pressure, $F=0.2$]{\includegraphics[height=.33\linewidth]{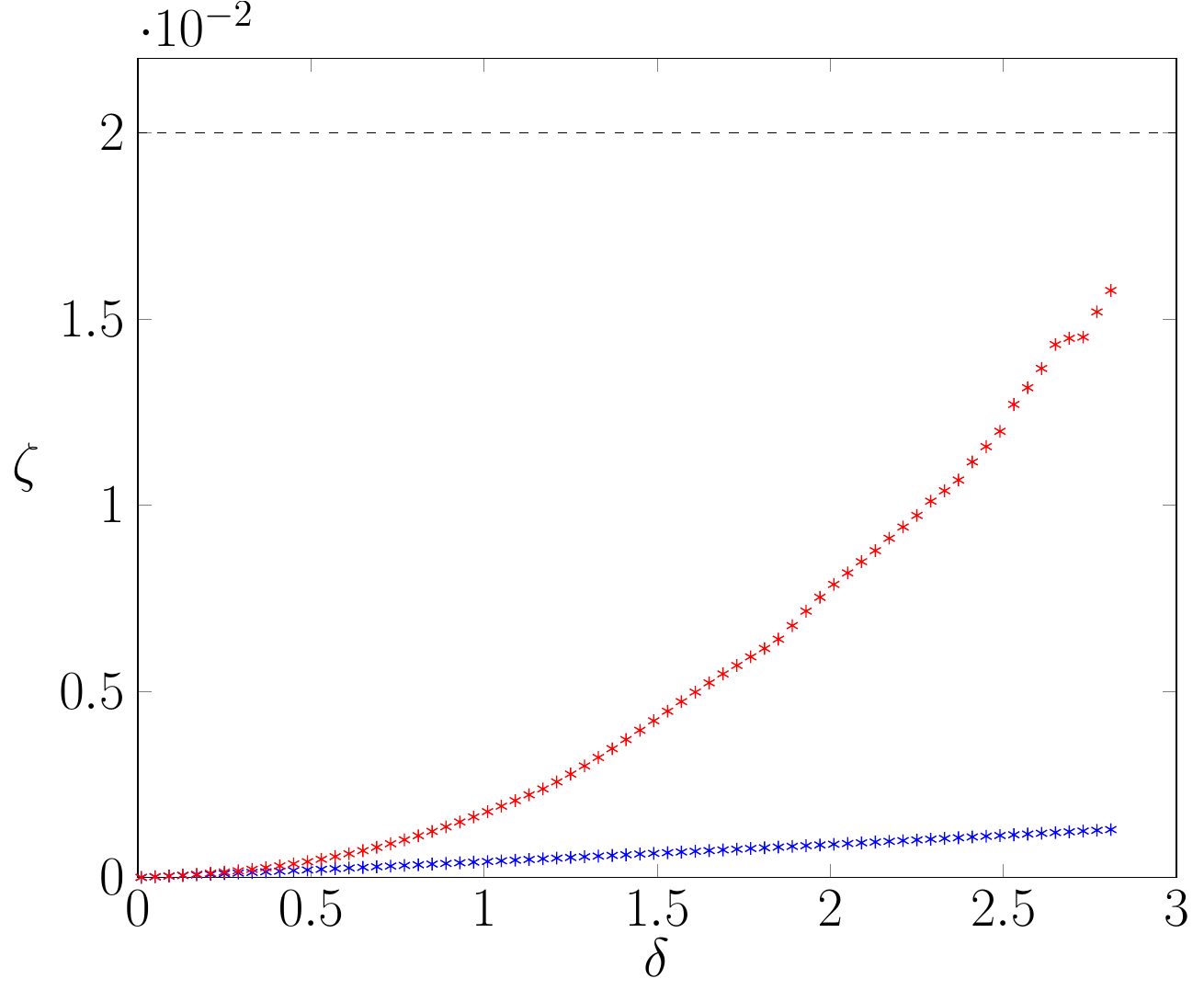}}\hspace{1.5ex}
		\subfloat[Pressure, $F=0.3$]{\includegraphics[height=.33\linewidth]{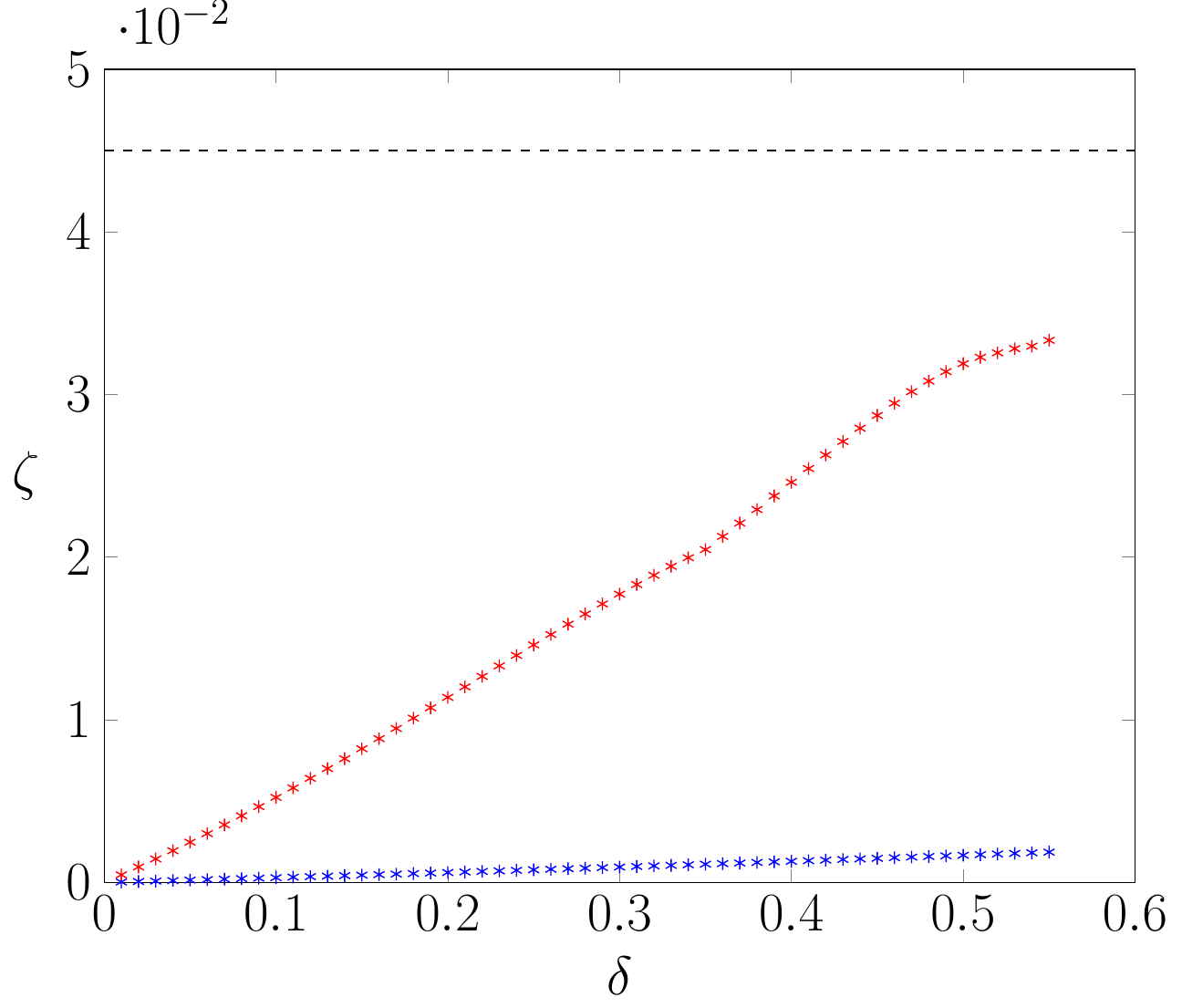}}
		\caption{A plot of the height of the initial peak (blue) and the first peak of the wave train (red) against the key parameter ($\epsilon$, $\mu$ or $\delta$) for flow past (a,b) a submerged source, (c,d) a submerged doublet, and (e,f) an applied pressure distribution.  Plots in (a,c,e) are for $F=0.2$ while those in (b,d,f) are for $F=0.3$. The dashed line is the Stokes limiting height, $\zeta=F^2/2$.}\label{fig:CentrelinePeaks}
	\end{figure}

	It is interesting to also track the height of the first peak of the wave train as $\delta$ increases.  For $F=0.2$ we see in Figure~\ref{fig:CentrelinePeaks}(e) that this first peak appears to be increasing with $\delta$ in a way that suggests the solution will reach a limiting configuration when this maximum height reaches $\max(\zeta)$ as $\delta\rightarrow \delta_{\mathrm{max}}$.  On the other hand, for $F=0.3$ in Figure~\ref{fig:CentrelinePeaks}(f), the trend is not as convincing, and so it is not clear whether a further increase in $\delta$ will lead to the solution ultimately reaching a limiting configuration.
	

	\section{Discussion}\label{sec:discussion}
	
	We have provided a rather detailed study of three-dimensional steady free-surface flows past a disturbance in the regime in which the background flow is considered `slow' or, in other words, the Froude number is small.  This work is motivated in part by the great deal of recent attention devoted to the complementary limit of large Froude number.  The three configurations we have focussed on are flow past a submerged point source, flow past a submerged point doublet, and flow past an applied pressure distribution.  In each case, the typical wave patterns share characteristics of a ship wake or a wake created by a submerged vessel like a submarine.  Importantly, for small Froude number (slow background flow or slowly moving ship) it is normally observed that the wake is dominated by transverse waves while for large Froude numbers (fast background flow or fast moving ship) the wake is dominated by divergent waves.  By examining a number of different limiting regimes we have been able to shine light on various examples in which the wave patterns have properties that are not normally observed.
	
	While there are many results in this study, one key finding is that the apparent wake angle $\theta_\mathrm{app}$ for small Froude numbers appears to decrease as the Froude number decreases.  Applying the definition of apparent wake angle used by \citet{darmon14}, where an arbitrary percentage of the maximum height is used to define the boundary of the wave, we find that $\theta_\mathrm{app}$ scales like $F$ for linearised flow past a submerged source or doublet (see (\ref{eq:lowFroudeAsymptsd})) and like $F^2$ for flow past an applied pressure (see (\ref{eq:lowFroudeAsymptpressure})).  These results complement the large Froude number scaling for which the apparent wake angle $\theta_\mathrm{app}$ (defined by tracking the highest peak of the waves as in \citet{pethiyagoda14b}) is proportional to $1/F$.  Another striking result is how for highly nonlinear regimes with small Froude number the wave patterns appear to be dominated by divergent waves (for the cases of flow past a submerged doublet and applied pressure) when the usual observation is that low Froude number flows are dominated by transverse waves.  These highly nonlinear steady wave patterns are rather noteworthy and are worth further study, especially as they approach their limiting configuration with a stagnation point forming at the highest wave peak.
	
	As pointed out as early as \citet{dagan75}  (in the context of two-dimensional flows), the perturbation schemes that give rise to linear problems are not uniform if we take the additional limit $F\rightarrow 0$.  For the present study, the linear problems are for $\epsilon\ll 1$, $\mu\ll 1$ and $\delta\ll 1$, and so the small-Froude-number limits in \S\ref{sec:linear} are strictly for $\epsilon\ll F \ll 1$, $\mu\ll F \ll 1$ or $\delta\ll F\ll 1$.  The details of the corresponding interesting and challenging problem in exponential asymptotics are contained in \citet{lustri13} and the time-dependent analogue \citep{lustri14} which, along with \citet{lustri19}, are the only previous studies of three-dimensional ship waves that use exponential asymptotics in the limit $F\rightarrow 0$.  In terms of future work, one obvious open problem in exponential asymptotics is to perform the equivalent analysis for $F\rightarrow 0$ with $\epsilon$, $\mu$ or $\delta$ fixed.  We close by discussing the challenges involved in approaching such a problem.
	
	The full nonlinear problem in exponential asymptotics ($F\rightarrow 0$ with $\epsilon$ or $\mu$ fixed) is superficially similar to the linear regime analysis presented in \S\ref{sec:linear}; however, attempting to formulate the problem in such a way that the exponential asymptotic techniques of \cite{lustri13} may be applied proves to be a significant challenge.  In the linear regime, the flow region is contained within the half space $z \leq 0$, while for the full nonlinear problem (which is a free boundary problem) this is not true.  Typically, nonlinear free boundary problems in fluid dynamics are studied analytically by applying a mapping to the flow region which fixes the position of the surface. In two dimensions, this is often accomplished by expressing the flow in terms of a complex potential, and applying subsequent mappings in order to contain the flow behaviour within a convenient domain, such as \citep{binder06,chapman06,lustri12}.  For three-dimensional problems, such conformal mappings are not available.
	
	There are two techniques that are normally used in three dimensions to fix free surface positions; boundary-fixing transformations, and potential mappings. Boundary fixing transformations typically involve defining a new variable $\eta = \zeta(x,y) - z$, where $\zeta(x,y)$ is the free-surface position. In this case, the surface is fixed at $\eta = 0$; however, the position of the singularity is now unknown. Three-dimensional extensions of complex potential mappings, such as Clebsch potentials \citep{lamb16}, offer more promise; however, these methods introduce another free boundary into the problem, as it must be determined whether any particular streamline of the flow originated at the submerged source, or upstream.
	
	Once the appropriate mapped region is determined, we expect that any subsequent analysis will be significantly more complicated for the nonlinear case. In particular, the governing equation for $\chi$ will no longer be independent of the source type, as in (\ref{eq:SingulantEq}), and hence we will need to determine this quantity for each different submerged obstacle. The systems obtained by applying exponential asymptotics to highly nonlinear fluid flow problems are often very difficult, or even analytically intractable, and must be solved numerically; see, for example \citet{lustri13a}. Nonetheless, studying the resultant highly nonlinear systems can still provide significant insight into the behaviour of water waves, and in particular, the interactions between various wave classes, such as the different wave types seen in \cite{lustri13}.

	\section*{Acknowledgements}The authors acknowledge the support of the Australian Research Council via the Discovery Projects DP140100933 (SWM), DP180103260 (SWM, TJM and RP) and DP190191190 (CJL).  SWM and CJL would like to thank the Isaac Newton Institute for Mathematical Sciences, Cambridge, for support and hospitality during the programme Complex Analysis: Techniques, Applications and Computations where part of the work on this paper was undertaken.  This programme was supported by the EPSRC grant EP/R014604/1.  SWM is grateful for the generous support of the Simons Foundation who provided further financial support for his visit to the Isaac Newton Institute via a Simons Foundation Fellowship.  SWM and CJL thank Jonathan Chapman for many fruitful discussions on low Froude number flows.
	
	\bibliographystyle{jfm}

	\appendix
	
	{
	\section{Power law relation for a general pressure distribution}\label{sec:powerLawPress}
	In this section we will partially derive the power law relation for apparent wake angle of a general pressure distribution. Given a pressure distribution $p(x,y)$ and its associated Fourier transform $\tilde{p}(k,\psi)$, where $k$ is the wavenumber and $\psi$ is the angle of propagation, we can rewrite (\ref{eq:exactPressureLinearInfinite}) and (\ref{eq:acrestpressure}) in terms of the general pressure distribution,
	\begin{align}
	\zeta(x,y) = &-\delta F^2 p(x,y)+\frac{\delta F^2}{2\pi^2} \int\limits_{-\pi/2}^{\pi/2}\,\int\limits_{0}^{\infty}\frac{k^2\tilde{p}(k,\psi)\cos(k[|x|\cos\psi+y\sin\psi])}{k-k_0}\,\,\mathrm{d}k\,\,\mathrm{d}\psi\notag\\
	&-\frac{\delta F^2 H(x)}{\pi}\int_{-\infty}^{\infty}\xi^2 \tilde{p}(k(\lambda),\psi(\lambda))\sin(x\xi)\cos(y\xi \lambda)\,\,\mbox{d}\lambda,\label{eq:generalPressureLinearInfinite}
	\end{align}
	\begin{equation}
	\hat{a}^p_\mathrm{crest}(\theta)	=(\lambda_1(\theta)^2+1)^2\sqrt{\frac{\cos\theta+\lambda_1(\theta)\sin\theta}{\cos\theta+(2\lambda_1(\theta)^3 +3\lambda_1(\theta))\sin\theta}}
	\,\frac{|\tilde{p}(k(\lambda_1(\theta)),\psi(\lambda_1(\theta))|}{|\tilde{p}(1/F^2,0)|},
\label{eq:acrestgeneralpressure}
	\end{equation}
	where $k(\lambda)=(\lambda^2+1)/F^2$, $\psi(\lambda)=\tan^{-1}\lambda$, $\xi$ is given by (\ref{eq:xi}) and $\lambda_1(\theta)$ is given by (\ref{eqn:parts1}).  We take the logarithm of both sides of (\ref{eq:acrestgeneralpressure}) and, anticipating that the wakes will narrow for decreasing Froude number,  expand the results as a Taylor series of about $\theta=0$, noting $\lambda_1=-\theta+\mathcal{O}(\theta^3)$ as $\theta\rightarrow 0$ to find
	\begin{equation}
	\ln(\hat{a}^p_\mathrm{crest})=
-\frac{\tilde{p}_\psi}{\tilde{p}}\theta
+\left(3+\frac{\tilde{p}_k}{F^2\tilde{p}}+\frac{\tilde{p}_{\psi\psi}}{2\tilde{p}}
-\frac{\tilde{p}_\psi^2}{2\tilde{p}^2}\right)\theta^2+\mathcal{O}(\theta^3),
\label{eq:genPressTheta}
\end{equation}
where subscripts indicate partial differentiation and the functions $\tilde{p}$, $\tilde{p}_k$, $\tilde{p}_\psi$, and $\tilde{p}_{\psi\psi}$ are evaluated at $k=1/F^2$ and $\psi=0$.

From this point we consider a specific pressure distribution, set $\hat{a}^p_\mathrm{crest}=\alpha$, $\theta=\theta_{\mathrm{app}}$ and attempt to take the limit as $F\rightarrow 0$.  Clearly the resulting scaling law heavily depends on the form of the pressure in Fourier space and its derivatives.  An axisymmetric pressure distribution will have $\tilde{p}_\psi=0$, which will simplify the matter.  For example, for the Gaussian pressure (\ref{eq:pressure}), $\tilde{p}_k/\tilde{p}$ evaluated at $k=1/F^2$, $\psi=0$ is $-1/(2\pi^2F^2)$, which gives the $F^2$ scaling (\ref{eq:lowFroudeAsymptpressure}).  On the other hand, a decaying pressure $p(x,y)=(1+x^2+y^2)^{-1}$ has $\tilde{p}_k/\tilde{p}=K_1(k)/K_0(k)$, where $K_0$ and $K_1$ are modified Bessel functions of the second kind, which means that the apparent wake angle scales exactly as in (\ref{eq:lowFroudeAsymptsd}).  For the non-axisymmetric pressure distributions (\ref{eq:pressuresource}) and (\ref{eq:pressuredoublet}), the scaled derivatives with respect to $\psi$ (i.e. ${\tilde{p}_{\psi}}/{\tilde{p}}$ and ${\tilde{p}_{\psi\psi}}/{\tilde{p}}$) do not depend on the Froude number and so again we arrive at the scaling $\theta_{\mathrm{app}}=\mathcal{O}(F)$.  More complicated pressure distribution may lead to different scalings.

	\section{Asymptotic approximation of $\mathrm{Re}(\chi)$}\label{sec:chiApprox}
	In this section we will derive the asymptotic approximation of $\mathrm{Re}(\chi)$ for $r\rightarrow 0$ by converting (\ref{eq:u}) and (\ref{eq:sroot}) to polar coordinates and choosing the positive sign for (\ref{eq:u}),
	\begin{equation}
	\chi=\frac{s-r\cos\theta}{s(2+s^2)},\label{eq:appu}
	\end{equation}
	where $s$ is one of the four roots to the quartic
	\begin{equation}
	r^2s^4+4r\cos\theta s^3+(r^2+3r^2\sin^2\theta+4)s^2+4r\cos\theta s+(4r^2\sin^2\theta+4)=0.\label{eq:appSroot}
	\end{equation}
	We use the perturbation
	\[
	s = s_0 + \frac{s_1}{r}+\mathcal{O}\left(\frac{1}{r^2}\right),
	\]
	in (\ref{eq:appu}) and (\ref{eq:appSroot}) to give the expansion
	\begin{equation}
	\chi=-\frac{r\cos\theta}{s_0(2+s_0^2)}+\frac{s_0^4 + 2s_0^2 + (2+3s_0^2)s_1\cos\theta}{s_0^2(s_0^2 + 2)^2}+\mathcal{O}\left(\frac{1}{r}\right),
	\end{equation}
	where $s_0$ is a root of
	\begin{equation}
	s_0^4+(1+3\sin^2\theta)s_0^2+4\sin^2\theta=0,\label{eq:appS0root}
	\end{equation}
	and
	\begin{equation}
	s_1=-\frac{2\cos\theta(s_0^2 + 1)}{2s_0^2 - 3\cos^2\theta + 4}.\label{eq:appS1}
	\end{equation}
	The solution to (\ref{eq:appS0root}) we are interested in is
	\begin{equation}
	s_0=\frac{i}{2}\sqrt{8-6\cos^2\theta+2\cos\theta\sqrt{9\cos^2\theta-8}},\label{eq:appS0}
	\end{equation}
	noting that within the Kelvin angle, $\cos\theta>\sqrt{8}/3$, $s_0$ is purely imaginary and therefore $s_1$ is purely real. Thus, we have
	\begin{equation}
	\mathrm{Re}(\chi)=\frac{s_0^4 + 2s_0^2 + (2+3s_0^2)s_1\cos\theta}{s_0^2(s_0^2 + 2)^2}+\mathcal{O}\left(\frac{1}{r}\right).\label{eq:appuExpand}
	\end{equation}
	Substituting (\ref{eq:appS1}) and (\ref{eq:appS0}) into (\ref{eq:appuExpand}) and simplifying gives us our asymptotic approximation for $\mathrm{Re}(\chi)$ (\ref{eq:uExpand})-(\ref{eq:uApprox}).
}
\end{document}